\documentclass[sn-mathphys,Numbered]{sn-jnl}% Math and Physical Sciences 

\usepackage{graphicx}%
\usepackage{multirow}%
\usepackage{amsmath,amssymb,amsfonts}%
\usepackage{amsthm}%
\usepackage{mathrsfs}%
\usepackage[title]{appendix}%
\usepackage{xcolor}%
\usepackage{textcomp}%
\usepackage{manyfoot}%
\usepackage{booktabs}%
\usepackage{algorithm}%
\usepackage{algorithmicx}%
\usepackage{algpseudocode}%
\usepackage{listings}%
\usepackage{lineno}
\usepackage{rotating}
\linenumbers

\usepackage{hyperref}
\usepackage{lineno}
\usepackage{xspace}
\usepackage{upgreek}
\usepackage{multirow}
\usepackage{euscript}
\usepackage{colortbl}
\usepackage{float}
\usepackage{siunitx}
\usepackage[T1]{fontenc}
\usepackage{caption}
\usepackage{comment}
\usepackage[]{moresize}
\usepackage[normalem]{ulem}           % for strike-through
\usepackage{microtype}			% for better typography

\theoremstyle{thmstyleone}%

\theoremstyle{thmstyletwo}%
\theoremstyle{thmstylethree}%

\newcommand{\Aegis}{AE\={g}IS\xspace}

\usepackage{verbatimbox}
\usepackage{microtype}

\begin{document}
\nolinenumbers

\title{CIRCUS: an autonomous control system for antimatter, atomic and quantum physics experiments}

\author[1,2,3]{M.~Volponi$^{\ast}$}\email{marco.volponi@cern.ch}
\author[1,4]{S.~Huck$^{\ast}$}\email{saiva.huck@cern.ch}

\author[3,2]{R.~Caravita}
\author[5]{J.~Zielinski}
\author[5]{G.~Kornakov}
\author[5]{G.~Kasprowicz}
\author[5]{D.~Nowicka}
\author[1,6]{T.~Rauschendorfer}
\author[7]{B.~Rien\"{a}cker}
\author[8]{F.~Prelz}

\author[9]{M.~Auzins}
\author[10]{B.~Bergmann}
\author[10]{P.~Burian}
\author[3,2]{R.~S.~Brusa}
\author[11]{A.~Camper}
\author[8,12]{F.~Castelli}
\author[13]{R.~Ciury\l{}o}
\author[8,14]{G.~Consolati}
\author[1]{M.~Doser}
\author[1]{L.~T.~Gl\"{o}ggler}
\author[5]{\L.~Graczykowski}
\author[1]{M.~Grosbart}
\author[3,2]{F.~Guatieri}
\author[1,15]{N.~Gusakova}
\author[1]{F.~Gustafsson}
\author[1]{S.~Haider}
\author[5]{M.~Janik}
\author[1]{G.~Khatri}
\author[13]{\L.~K\l osowski}
\author[1,9]{V.~Krumins}
\author[5]{L.~Lappo}
\author[13]{A.~Linek}
\author[1,11]{J.~Malamant}
\author[2,3]{S.~Mariazzi}
\author[3,2]{L.~Penasa}
\author[16]{V.~Petracek}
\author[13]{M.~Piwi\'nski}
\author[10]{S.~Pospisil}
\author[2,3]{L.~Povolo}
\author[17]{S.~Rangwala}
\author[7]{B.~S.~Rawat}
\author[7]{V.~Rodin}
\author[11]{O.~M.~R{\o}hne}
\author[11]{H.~Sandaker}
\author[10]{P.~Smolyanskiy}
\author[18]{T.~Sowi\'nski}
\author[5]{D.~Tefelski}
\author[1]{T.~Vafeiadis}
\author[7]{C.~P.~Welsch}
\author[1]{T.~Wolz}
\author[13]{M.~Zawada}
\author[19,20]{N.~Zurlo}

\affil[1]{Physics Department, CERN, 1211~Geneva~23, Switzerland}
\affil[2]{TIFPA/INFN Trento, via Sommarive 14, 38123~Povo, Trento, Italy}
\affil[3]{Department of Physics, University of Trento, via Sommarive 14, 38123~Povo, Trento, Italy}
\affil[4]{Institute for Experimental Physics, Universit\"{a}t Hamburg, 22607 Hamburg, Germany}
\affil[5]{Warsaw University of Technology, Faculty of Physics ul. Koszykowa 75, 00-662, Warsaw, Poland}
\affil[6]{Felix Bloch Institute for Solid State Physics, Universit\"{a}t Leipzig, 04103 Leipzig, Germany}
\affil[7]{University of Liverpool, UK and The Cockcroft Institute, Daresbury, UK}
\affil[8]{INFN Milano, via Celoria 16, 20133~Milano, Italy}
\affil[9]{University of Latvia, Department of Physics Raina boulevard 19, LV-1586, Riga, Latvia}
\affil[10]{Institute of Experimental and Applied Physics, Czech Technical University in Prague, Husova 240/5, 110 00, Prague 1, Czech Republic}
\affil[11]{Department of Physics, University of Oslo, Sem Sælandsvei 24, 0371~Oslo, Norway}
\affil[12]{Department of Physics ``Aldo Pontremoli'', University of Milano, via Celoria 16, 20133~Milano, Italy}
\affil[13]{Institute of Physics, Faculty of Physics, Astronomy, and Informatics, Nicolaus Copernicus University in Toru\'n, Grudziadzka 5, 87-100 Toru\'n, Poland}
\affil[14]{Department of Aerospace Science and Technology, Politecnico di Milano, via La Masa 34, 20156~Milano, Italy}
\affil[15]{Department of Physics, NTNU, Norwegian University of Science and Technology, Trondheim, Norway}
\affil[16]{Czech Technical University, Prague, Brehov\'a 7, 11519~Prague~1, Czech Republic}
\affil[17]{Raman Research Institute, C. V. Raman Avenue, Sadashivanagar, Bangalore 560080, India}
\affil[18]{Institute of Physics, Polish Academy of Sciences, Aleja Lotnikow 32/46, PL-02668 Warsaw, Poland}
\affil[19]{INFN Pavia, via Bassi 6, 27100~Pavia, Italy}
\affil[20]{Department of Civil, Environmental, Architectural Engineering and Mathematics, University of Brescia, via Branze 43, 25123~Brescia, Italy}

\abstract{
A powerful and robust control system is a crucial, often neglected, pillar of any modern, complex physics experiment that requires the management of a multitude of different devices and their precise time synchronisation.
The \Aegis collaboration presents CIRCUS, a novel, autonomous control system optimised for time-critical experiments such as those at CERN’s Antiproton Decelerator and, more broadly, in atomic and quantum physics research. Its setup is based on Sinara/ARTIQ and TALOS, integrating the ALPACA analysis pipeline, the last two developed entirely in \Aegis. It is suitable for strict synchronicity requirements and repeatable, automated operation of experiments, culminating in autonomous parameter optimisation via feedback from real-time data analysis.
CIRCUS has been successfully deployed and tested in \Aegis; being experiment-agnostic and released open-source, other experiments can leverage its capabilities.
}


\maketitle

%%%%%%%%%%%%%%%%%%%%%%%%%%%%%%%%%%%%%%%%%%%%%%%%%%

% Introduction

%%%%%%%%%%%%%%%%%%%%%%%%%%%%%%%%%%%%%%%%%%%%%%%%%%

\section{Introduction}
\label{section_intro}

Control systems are, generally speaking, combinations of hardware and software with the ability to modify the operation and/or configuration of other elements of a system and are in charge of the management of that system. {\it Autonomous} control systems are such that can operate with little to no human supervision. They are applied in any imaginable field, from satellites to dishwashers. Control systems for nuclear, atomic, and quantum physics experiments are a special category because they need to deal with systems that are continuously upgraded, fixed, and reshaped. For this reason, they need to maintain stability, reliability and reproducibility while allowing for the flexibility necessary for the experiment to mutate\footnote{This is different from the demands of the control systems of big observational experiments (such as LHC main experiments, or neutrino telescopes), which are less prone to change.}. The nature of these experiments puts a range of constraints on the control system: nanosecond-precise execution, multiple computer synchronisation, interfacing with different hardware using multiple interfaces, and easy extendability, among others.

The experiments at CERN's Antiproton Decelerator (AD) complex~\cite{aegis_proposal:07, alpha_proposal, base_proposal, asacusa_proposal, gbar_proposal, puma:19}, which investigate the asymmetry between matter and antimatter in the universe, are examples for such experiments. They rely on the combination of techniques from photonics, plasma, quantum, nuclear, and particle physics. For example, to be able to manipulate antimatter, it has to be isolated from ordinary matter to avoid annihilation. Antiprotons are typically trapped in ultra-high vacuum inside electromagnetic traps in the form of non-neutral plasmas~\cite{Gabrielse86, madsen_houches:14}, often sympathetically cooled and manipulated using electrons~\cite{Gabrielse1989, aegis_compression:18}. In combination with cold positron plasmas \cite{surko_rev:15}, they are used to form antihydrogen \cite{athena_nat:02, atrap_hbar:02}, which can be trapped \cite{alpha_nat:11} and probed using techniques such as spectroscopy~\cite{alpha_1s2s:16}. Manipulation and preparation of specific quantum states of anti-atoms is currently also explored in several experiments~\cite{asacusa_beam:14}.

One of these experiments is \Aegis (Antimatter Experiment: Gravity, Interferometry, Spectroscopy)~\cite{aegis_elena:18}, whose main aim is to measure the gravitational displacement of a horizontal pulsed beam of antihydrogen ($\rm{\overline{H}}$) using a moiré deflectometer~\cite{aegis_natc:14}. The experiment has developed a unique pulsed scheme which is able to provide precise knowledge of the $\rm{\overline{H}}$ formation time, control the final antihydrogen temperature, and manipulate its excitation state, among others. The formation of antihydrogen is based on the charge-exchange reaction between Rydberg-excited positronium (Ps) atoms and trapped, cold antiprotons from the CERN decelerators~\cite{CHARLTON1990143,krasnicky_pra:16}. The \Aegis apparatus~\cite{aegis_apparatus:15} comprises two cylindrical cryostats containing superconducting magnets of \SI{5}{\tesla} and a \SI{1}{\tesla}, respectively. A Penning-Malmberg trap in the \SI{5}{\tesla} region is optimised for trapping and cooling antiprotons, while a second trap in the \SI{1}{\tesla} region is used to form antihydrogen. The axial confinement of charged particles is achieved by the more than 60 electrodes forming the two traps and, to minimise the losses of trapped antiprotons, an ultra-high vacuum of $10^{-13}$ \si{\milli\bar} or better is maintained. Additionally to the electrodes, the manipulation of the accumulated particle plasmas and anti-atoms is done with a set of q-switched pulsed lasers, relevant for the excitation of positronium to efficiently produce $\rm{\overline{H}}$. The apparatus is equipped with a Micro-Channel Plate (MCP) detector at the end of the two cryostats, a two-layer scintillator fibre tracker for detecting the annihilation~\cite{aegis_fact:13,amsler_cryogenic_2020}, plastic scintillators~\cite{Zurlo:2020grr}, and an optical fibre bundle to monitor the light from the lasers. The entry region of the antiproton beam from the AD also serves to bring in positrons from an accumulator, which are then converted to positronium in a dedicated silicon nano-channel target~\cite{mariazzi:21,mariazzi_positronium_2010,mariazzi_high_2010}. The complexity of the apparatus gives the possibility to investigate different phenomena: for example, attempts to laser-cool positronium atoms are currently ongoing, using the experience of positronium generation and the recently upgraded laser system. The installation of an additional trap for heavy ion generation is also ongoing, which will enable \Aegis to perform studies on the formation processes of highly-charged antiprotonic heavy ions.

In the initial phase of the experiment, sequences of operations pre-defined by the users and executed by monolithic control systems developed progressively over the years, on top of a custom-made electronics system with \si{\nano\second} synchronisation capabilities, were adequate to successfully produce antihydrogen atoms in a pulsed modality~\cite{aegis_hbar:21}. In the process of establishing antihydrogen formation, however, the limits of this approach became evident: the lack of programming structures to tackle the increasing complexity of experimental sequences; the need of online procedure debugging capabilities; the limited re-usability of the written sequences. In other words, the necessity of an end-user interface providing the features of a standard programming language emerged, although still requiring arbitrary waveform generation and \si{\nano\second} synchronisation capabilities to allow complex non-neutral plasma manipulations \cite{aegis_compression:18} as well as Ps formation and laser excitation~\cite{aegis_neq3:16}. 

In fact, as often occurring in complex experiments, the software infrastructure of the \Aegis apparatus consisted of multiple independent subsystems (e.g. antiproton trap, positron apparatus, laser systems, detectors, etc.), managed by a set of computers running several control programs, all independently written and connected by pre-defined interfaces, which in turn had to be adapted to the changing needs of the experiment. While, with this approach, each single subsystem could be initially developed independently from the others, the performance of coordinated experiments (like antihydrogen production) required a significant human effort to operate the entire system as a whole, as the individual control programs needed constant monitoring during the data-taking periods.

Different examples of control systems for physics experiments exist~\cite{perego_scalable_2018,starkey_scripted_2013,agraz_labview-based_2014,trenkwalder_flexible_2021,keshet_distributed_2012}, which share most of the concepts expressed above and propose different solutions to the aforementioned problems. Nevertheless, the interfacing capability is often limited, and, furthermore, none of them is envisaged with automation as the main driving force: the possibility of letting a control system run in full autonomy, especially with a feedback loop based on acquired data, relies on layers of self checks and self consistency, which are not straightforward to implement.

Furthermore, the size and complexity of experiments like \Aegis renders impossible the entire control to be performed by a real-time code residing on a Field-Programmable Gate Array (FPGA). The multitude of interfaces required by the different instruments and the diverse time scales (nanoseconds for time-critical operations, minutes for an entire measurement sequence) cannot be provided by such a solution.

For these reasons, the \Aegis collaboration has designed a generalised experiment control system that is customisable to individual experiments' specific requirements. This flexibility benefits the \Aegis experiment (as it allows it to evolve smoothly to changing requirements), but equally importantly, the system was constructed with the needs of the much wider atomic and quantum physics community in mind. This control system incorporates a programmable end-user interface, providing advanced synchronisation, watchdog, error management, and online decision making features, re-enforced by an active feedback loop from the acquired data. This re-design specifically targeted complexity reduction of experimental procedures by standardising established sub-procedures into libraries, and by increasing stability, reliability, and autonomy. With this as the baseline, the subsequent implementation of increasing layers of automation and autonomy becomes feasible, strengthening the trust in the system by cycles of campaigns of implementation and debugging. 

The implemented solution merges the capabilities of a real-time code with a distributed slow-control system that unifies the computer in a single entity and brings together all the features described above, so as to partially remove the operators' need to control the running procedures. The control system itself is completely experiment-agnostic (technically, it could be used to control experiments outside the realm of physics as well), and it is released open-source so that other experiments can profit from the effort.

The high level of automation is a choice motivated further by  the upgrade of the AD to the new Extra Low ENergy Antiproton (ELENA) ring~\cite{bartmann_progress_2013}. ELENA is a small synchrotron with a \SI{30}{\meter} circumference, used to further decelerate AD antiprotons from \SI{5.3}{\MeV} down to \SI{100}{\keV} and finally transfer them to the experiments present at the AD. This results in an increase of one to two orders of magnitude in the trapping capabilities of the experiments. With ELENA, the operation scheme and the share of the $\rm{\overline{p}}$ beam has changed from experiment-specific allocated time slots of 8 hours to shared access and continuous 24/7 operation, increasing the shift personnel needs by a factor of three.

In this article, the new control system, called CIRCUS Computer Interface for Reliably Controlling, in an Unsupervised manner, Scientific experiments), is presented, with the specific implementation in the \Aegis experiment given as an example. It was designed around the Sinara/ARTIQ open hardware/software platform~\cite{sinara:20,artiq:16}, embedded within a LabVIEW\texttrademark~\cite{bitter2006labview}-based control framework called TALOS, providing the asynchronous high-level functionalities. The creation of experimental hardware procedures is done in the ARTIQ programming language (based on Python), which allows for \si{\nano\second}-synchronous operation scheduling on the Sinara hardware. The new control system has been used in \Aegis antiproton campaigns with ELENA and proved to be autonomous and reliable, while facilitating fast development of experimental procedures with version control, structured debugging, and agile development. 

% structure of the paper 
 The article is structured as follows: general requirements imposed by scientific goals are outlined in Sec.~\ref{section_requirements}. The new electronics setup is depicted in Sec.~\ref{section_hardware}, explaining the functionalities of the Sinara ecosystem. The overall software control system is then introduced in the following Sec.~\ref{section_software}, encompassing ARTIQ, the library for programming Sinara, and TALOS, the modular distributed slow-control system. An overview of the \Aegis Data Acquisition System is offered in Sec.~\ref{section_DAQ}, as an example case. Similarly, in Sec.~\ref{section_ALPACA}, the online and offline analysis system is shown, with feedback capabilities on the control system; a successful example of this application is described in in Sec.~\ref{section_automation_laser}. Subsequently, the CIRCUS control system validation is presented in Sec.~\ref{section_results}. Last, the performance of the new setup is evaluated and foreseen future implementations are outlined.

%%%%%%%%%%%%%%%%%%%%%%%%%%%%%%%%%%%%%%%%%%%%%%%%%%

% Methods

%%%%%%%%%%%%%%%%%%%%%%%%%%%%%%%%%%%%%%%%%%%%%%%%%%

\section{Methods}

%%%%%%%%%%%%%%%%%%%%%%%%%%%%%%%%%%%%%%%%%%%%%%%%%%

% Requirements

%%%%%%%%%%%%%%%%%%%%%%%%%%%%%%%%%%%%%%%%%%%%%%%%%%

\subsection{Requirements for the autonomous control system} 
\label{section_requirements}

The design of the control system is driven by the requirements that this class of experiments has. A review of the literature was performed, to take some examples of atomic and quantum experiments~\cite{chouFrequencyComparisonTwo2010, gonzalezImprovedNeutronLifetime2021, hinkleyAtomicClock102013, millenQuantumExperimentsMicroscale2020, thomasEntanglementDistantMacroscopic2021}, and relate their requirements to the ones derived from the experience of realising the first pulsed source of antihydrogen in \Aegis~\cite{aegis_hbar:21}. 
The comparison showed that this class of experiments share similar requirements, which can be subdivided into four categories: interface requirements with the particle source; trap operations; particle and laser synchronisation; general slow control, data acquisition (DAQ) and networking. 

Therefore, we decided to use the \Aegis requirements as a base for the design of the control system: the generality of these requirements renders a system satisfying them applicable to a broad range of tasks. In the following, their rationale is exposed, and the requirements are then summarised in Table~\ref{reqs_table}.

\emph{Requirements of the particle source interface:} \Aegis obtains the antiprotons in bunches from the AD--ELENA complex. Consequently, the experiment is synchronised to the decelerator stack via a set of hardware triggers occurring at different times during each $ \approx \SI{120}{\second} $ antiproton cycle: the AD injection trigger (occurring at the beginning of the cycle), the AD extraction/ELENA injection trigger (occurring $ \approx \SI{20}{\second}$ before antiproton delivery), a bunch pre-arrival trigger (occurring $ \approx \SI{20}{\micro\second} $ before antiproton extraction from ELENA) and a bunch arrival trigger (synchronous with the extraction from ELENA). The bunch is approximately \SI{150}{ns} (FWHM) long. Antiprotons are delivered to the experiment at \SI{100}{\keV} energy, which is further reduced by a thin foil (ca.~\SI{1500}{\nano\meter} of kapton) to about \SI{10}{\kilo\electronvolt}. Antiprotons are subsequently caught by means of a pulsed high-voltage Malmberg-Penning trap operated up to \SI{15}{\kilo\volt} in a \SI{5}{\tesla} magnetic field. The timing of the trap has to be fine-tuned in $ \approx \SI{10}{\nano\second} $ steps. 

\emph{Requirements for trapped particle manipulations:} a typical antihydrogen production sequence involves several manipulations steps of trapped particles (in the form of non-neutral plasmas), performed with low-voltage electrodes of the Malmberg-Penning trap in the \SI{1}{\tesla} region. These have to be controlled in the $ \pm \SI{200}{\volt} $ range, by arbitrary function generators. 
An accuracy of \SI{10}{\milli\volt} or better is needed to allow for accurate potential ramps and thus enable measurements of the plasma space charge and temperature \cite{Beck1992} as well as evaporative \cite{Andresen2010a} and adiabatic cooling \cite{Gabrielse2011}. Standard manipulations in traps include both slow (several seconds) and fast (less than a millisecond) ramps, fast extraction of particles with $ \approx \SI{100}{\nano\second} $ ($ \approx \SI{100}{\micro\second}$) pulses for electron (antiproton) extraction respectively, as well as application of radiofrequencies (RF) in the $\SI{1}{\kilo\hertz}-\SI{100}{\mega\hertz}$ range for plasma heating or cooling and density control with the Rotating Wall technique \cite{aegis_compression:18}. Often, these procedures are combined, and the ability to synchronise events with the accuracy of \SI{1}{\nano\second} over several hours is required.

\emph{Requirements of particle and laser synchronisation:} antihydrogen formation via charge-exchange reactions with trapped antiprotons requires the control of the times of positronium formation and laser excitation to its Rydberg levels at the \SI{}{\nano\second} accuracy level, as well as triggering the diagnostic scintillation and Micro-Channel Plate (MCP) detectors, as detailed in \cite{aegis_neq3:16,aegis_velocimetry:19}. This is due to the fact that the excitation laser has to be carefully synchronised according to its beam shape and position to obtain efficient positronium excitation. Hardware trigger lines allowing time adjustment of $\SI{1}{\nano\second}$ or better and jitters of $ < \SI{0.5}{\nano\second}$ are required. 

\emph{Slow control, DAQ and networking requirements:} these include all the procedural sequences of trap initialisation, synchronisation on slow scales, computer responsiveness, data upload to the Data Acquisition System, etc., which admit a considerable jitter between the moment the command is issued and its execution, and must not interfere with the experimental sequence (typically in the order of \SI{100}{\milli\second}). Network communication has to guarantee a smooth control flow: the communication speed among the various machines needs to be at least an order of magnitude faster than the timescale of PC operations. 

\begin{table}[htbp]
    \centering
    \begin{tabular}{lr}
        \multicolumn{2}{c}{\bf{Particle source interfacing}} \\
        \hline \hline       
        High-voltage catching potential range & 0--15 {\si{\kilo\volt}} \\
        High-voltage potential accuracy & < \SI{10}{\volt} \\
        High-voltage gate raising edge duration & < 100\,{\si{\nano\second}} \\
        High-voltage gate temporal tuning accuracy  & < 10\,{\si{\nano\second}} \\
        \hline
        AD injection trigger synchronization & < 5\,s \\
        ELENA injection trigger synchronization & < 1\,s \\
        Bunch pre-arrival trigger synchronization & < 1\,{\si{\micro\second}} \\
        Bunch arrival trigger synchronization & < 10\,{\si{\nano\second}} \\
        \\
        \multicolumn{2}{c}{\bf{Trapped particle manipulation}} \\
        \hline \hline
        Low-voltage potential range & $\pm$ 200 \,{\si{\volt}} \\        
        Low-voltage potential ramps duration & \SI{100}{\micro\second} - \SI{10}{\second} \\        
        Maximum absolute calibration error & < 5\,{\si{\milli\volt}} \\
        Maximum noise band amplitude & < 1 \,${\si{\milli\volt}}_\mathrm{rms}$ \\          
        \hline  
        Fast pulse settling time & < 30\,{\si{\nano\second}} \\
        Fast pulse duration range & \SI{100}{\nano\second} - \SI{100}{\micro\second} \\
        Fast pulse timing adjustment & < 10\,{\si{\nano\second}} \\
        Fast pulse timing jitter & < 1\,{\si{\nano\second}} \\
        \hline
        RF frequency range & \SI{1}{\kilo\hertz} - \SI{100}{\mega\hertz} \\
        RF amplitude range & \SI{10}{\milli\volt} - \SI{5}{\volt} \\
        \\
        \multicolumn{2}{c}{\bf{Particle and laser synchronisation}} \\
        \hline \hline
        Positron/laser triggers time adjustment & < 1\,{\si{\nano\second}} \\        
        Positron/laser trigger jitter & < 0.5\,{\si{\nano\second}} \\
        Detector arming timing accuracy & < 100\,{\si{\milli\second}}\\
        Detector triggering timing accuracy & < 10\,{\si{\nano\second}}\\
        \\
        \multicolumn{2}{c}{\bf{Slow control, DAQ and networking}} \\
        \hline \hline
        PC--PC message delay & < 100\,{\si{\milli\second}}\\
        Real-time--PC message delay & < 10\,{\si{\milli\second}} \\
        DAQ run start/stop time & < 10 \, \si{\second} \\
        Data availability for online analysis & < 5 \, \si{\second}
    \end{tabular}
    \caption{Summary of the different technical requirements set on the control system from experiments needs.}
    \label{reqs_table}
\end{table}

%%%%%%%%%%%%%%%%%%%%%%%%%%%%%%%%%%%%%%%%%%%%%%%%%%%%%

% Control System Hardware

%%%%%%%%%%%%%%%%%%%%%%%%%%%%%%%%%%%%%%%%%%%%%%%%%%%%%

\subsection{The control system hardware}
\label{section_hardware}

For atomic and quantum physics experiments, the necessity to operate (parts of) the measurements with ns-precision is fundamental (as seen in~\ref{section_requirements}). Hence, the control system electronics have a pivoting role in reaching the scientific objectives.

In \Aegis, the main components of the control system electronics belong to the Sinara \cite{sinara:20} ecosystem. Sinara features a versatile, open-source hardware portfolio which was originally developed for quantum information experiments utilising the ARTIQ control software \cite{artiq:16} (see section \ref{section_artiq}). The Sinara hardware provides compact, modular, reproducible and reliable electronics capable of controlling intricate, time-critical experiments. It is particularly optimised for experimental setups which are limited in space, as is the case inside the AD, and, thanks to its standardised and modular nature, assures the long-term maintainability of the control system.

While Sinara was chosen for the above reasons and is easily applicable to a multitude of very different procedures in quantum and atomic physics experiments, ARTIQ can be used in combination with hardware and peripherals from other manufacturers capable of nanosecond timing as well, if controlled by a dedicated FPGA.

\begin{center}
\begin{figure}[htpb]
\centering
\includegraphics[width=\textwidth]{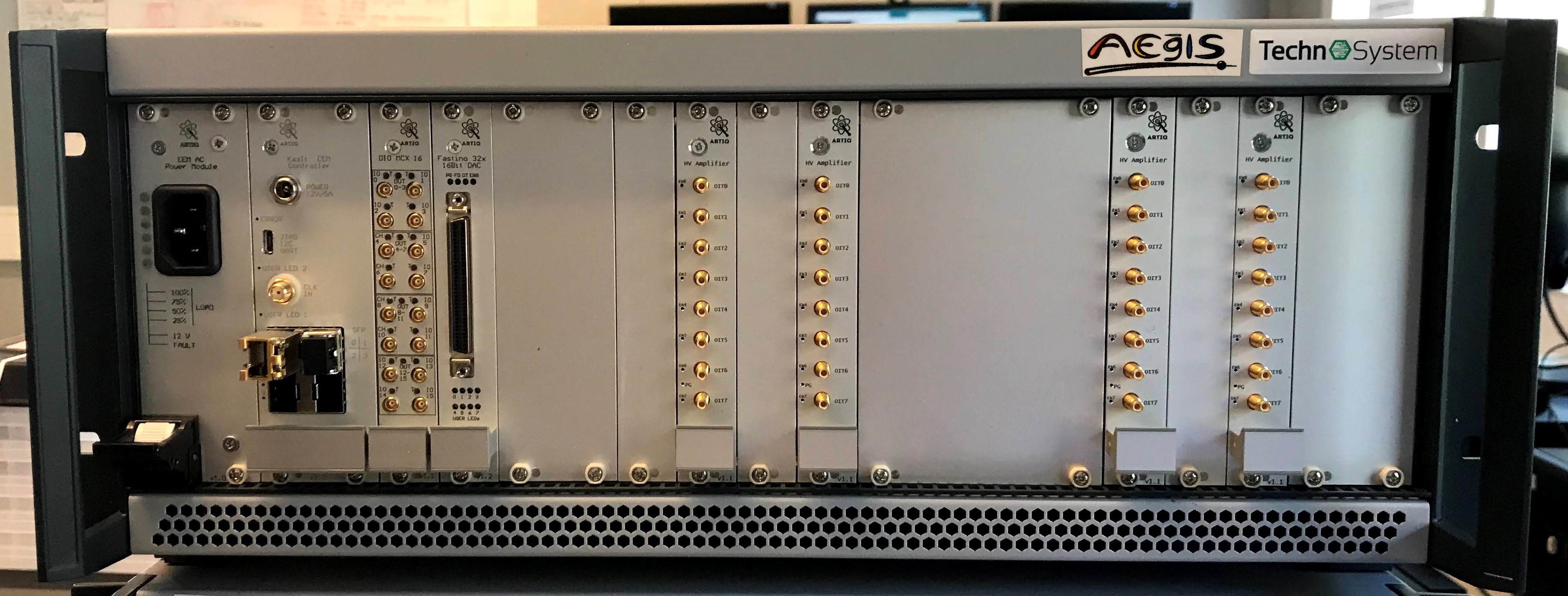}
\caption{Photograph of one of three fully equipped Sinara electronics crates of the \Aegis trap control system, including (from left to right) power module, Kasli carrier, digital I/O units, Fastino DAC, and four high-voltage amplifier boards.}
\label{crateandrack}
\end{figure}
\end{center}

As shown in Fig.~\ref{crateandrack}, the hardware of the \Aegis trap control system is organised in three rack-standard Eurocard 84 HP electronics crates with dimensions of 50\,x\,20\,x\,35\,cm, which allow to connect a variety of modules.

The main controller is called Kasli (see Fig.~\ref{crateandrack}). It comprises an Artix-7 Field-Programmable Gate Array (FPGA) and can be used as a stand-alone core device or in combination with additional carriers as a repeater or satellite of DRTIO (Distributed Real Time Input/Output) communication through optical fibre links, facilitating a stable, high-speed Gbps transfer of time and data information between the devices. This second option allows for a fast propagation of both a clock signal (internally generated or externally connected) and the control communication between controllers, thus offering straightforward adaptations and extensions of the experiment. Software communication with the Sinara electronics is facilitated via Kasli’s high-speed Gigabit Ethernet port.
Each Kasli is capable of controlling up to twelve extension modules with various purposes.

Each Sinara crate used in \Aegis contains a Kasli carrier combined with digital I/O units and fast DAC modules, called Fastino, from the Sinara repertoire, as well as \SI{1}{\MHz} high-voltage amplifiers, which have been custom-designed for the requirements of the \Aegis experiment.

The digital I/O cards are used for the reception and provision of high-speed \SI{}{\ns} TTL trigger signals between the sub-systems of the entire experimental setup. 16 MCX connectors are compactly arranged on each single, thin module and their direction of input or output can be individually configured in batches of four.

Each Fastino provides simultaneous \SI{3}{\mega\bit\per\second} digital-to-analog conversion for 32 channels, yielding stable output voltages in the range of \SI{\pm10}{\volt} with a \SI{16}{\bit} resolution. The Fastino DAC channels can either be used directly to supply low voltages in this range or be connected in batches of eight to the high-voltage amplifier modules.

One such amplifier unit comprises eight channels, each of which is capable of a 20-fold amplification of the output voltage of one Fastino channel respectively, i.e. yielding voltages of up to \SI{\pm200}{\volt}. The high-voltage amplifiers are equipped with individual OptoMOS\textsuperscript{\textregistered} relays, allowing to isolate the outputs and prevent the noise from the amplifiers from propagating to the connected load.
The main control electronics of the \Aegis setup are formed by three of the described Sinara crates: two (one Kasli acting as DRTIO repeater, the other as satellite) provide the high-voltage output channels for the synchronous potential steering of the electrodes of the \SI{5}{\tesla} trap region of the experiment, required for antiproton capture and electron cooling, while the third crate is used for the control of the \SI{1}{\tesla} antihydrogen production trap electrodes.

During the ELENA/AD antiproton run campaigns, the fast digital I/O units have demonstrated reliable acquisition and processing of the incoming trigger signals, essentially enabling the steering of the trap potentials with the required timing for the capture of antiprotons.

\begin{figure}[H]
\begin{center}
\includegraphics[width=0.8\textwidth]{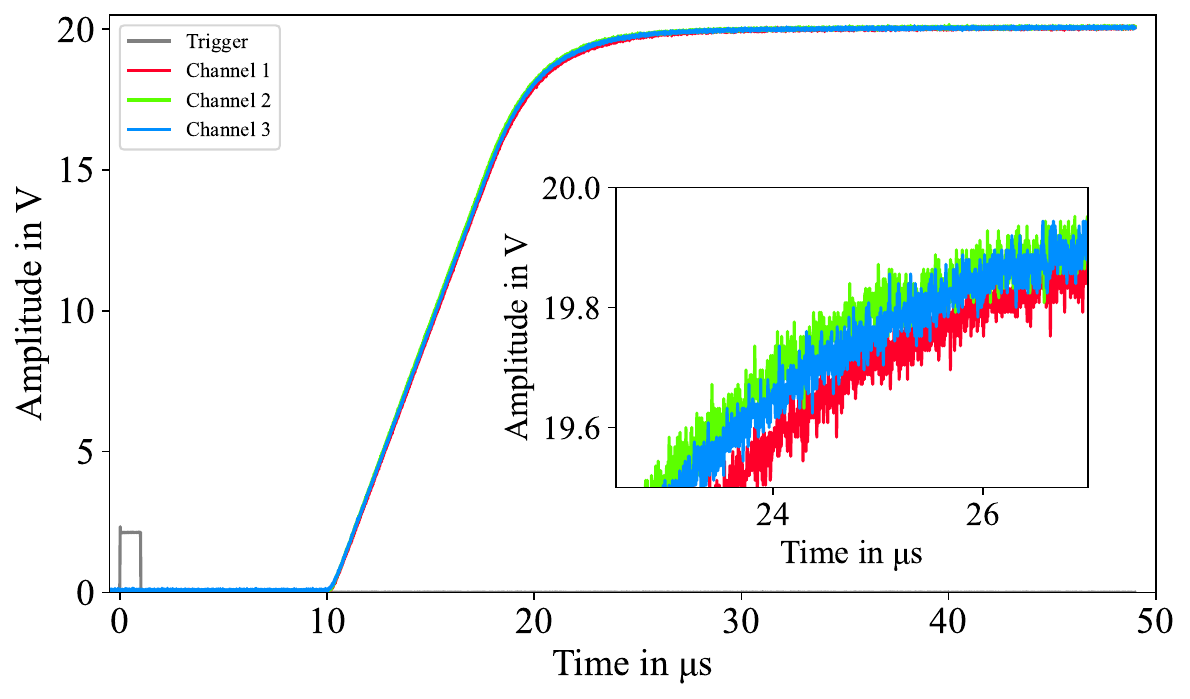}
\caption{Synchronous voltage ramp-up to \SI{20}{\volt} on three high-voltage amplifier channels \SI{10}{\us} subsequent to the arrival of a common trigger pulse at zero time in the figure. The inset shows a zoom to the shoulder region for a better visualisation of the synchronicity.}
\label{plots}
\end{center}
\end{figure}

In Fig.~\ref{plots}, examples of output signals of three HV amplifier channels are shown. They are produced by sending an external trigger pulse to the digital I/O unit and subsequently setting a voltage on three of the Fastino channels. The voltage is amplified by the connected amplifier units. The final output is recorded using an oscilloscope and read out via LabVIEW\texttrademark. The Sinara system is thus found to be able to satisfy the timing requirements of the \Aegis experiment: reactions to triggers on the microsecond scale and synchronous control of the output channel voltages are reliably provided.
All amplifier channels have been calibrated individually together with their corresponding Fastino DAC channels to reliably provide the required voltage despite their different offset and voltage precision step values. With this calibration, each channel voltage can be set with an accuracy of few \SI{}{\milli\volt}, which is comparable to the precision of the \SI{6}{\milli\volt} steps generated by the \SI{16}{\bit} nature of the Fastino. The calibration procedure is described in Appendix \ref{appendix_calibration}.

In addition to the electronics controlling the trap system and providing inter-system triggers, two additional Sinara crates have been successfully commissioned to run the laser system and provide synchronisation between the two involved lasers despite their difference in frequency (see section \ref{section_results}). In order to be able to do this, the \SI{1}{\tesla} Kasli, in addition to controlling the respective trap system, is used as the master core for two satellite Kasli devices, both of which control a digital I/O card with BNC output for triggering the sequences needed for laser operation\footnote{The BNC digital I/O units work in the same way as the MCX units except for comprising only eight channels instead of 16.}. Furthermore, the new control electronics have been successfully integrated as part of the \Aegis positron system to provide triggers for the positron preparation sequence and synchronize it to the rest of the experiment. Further extensions of the control infrastructure, e.g. dedicated Sinara crates for the positron system and to future-proof the use of the Rotating Wall technique for plasma compression, are ongoing.

The Sinara hardware is a central component in the new \Aegis control system, which drives all integral parts of the experiment. The software will be presented in the following section.

There are two relevant additional electronics components, which have been integrated in the new control system setup and are fully steerable programmatically.
The first is a pulser device which provides ns-synchronised pulses of variable length to the electrodes, with tunable amplitude in the voltage range  provided by the Sinara Fastino plus amplifier channels. The trigger signals for this pulser are given by the Sinara digital I/O units, while the amplitude is determined by internal DAC units. The inclusion of this functionality is vital for the efficient and fast transport of particles between the different trapping regions inside the experiment.
The second component is a waveform synthesizer with multiple channels, which can be used to add phase-shifted sinusoidal signals of up to \SI{5}{\volt} in a frequency range of 0 to \SI{30}{\mega\hertz} to the so-called sectorised electrodes. These electrodes are separated into four sectors around their centre, i.e. around the central axis of the trap. By applying the sinusoidal signals with a phase shift between the four sectors, it is possible to employ the Rotating Wall (RW) technique for a manipulation of the radial dimension of the contained particle plasma. This component is also currently operated by the new control system.

%%%%%%%%%%%%%%%%%%%%%%%%%%%%%%%%%%%%%%%%%%%%%%%%%%%

% Control system software

%%%%%%%%%%%%%%%%%%%%%%%%%%%%%%%%%%%%%%%%%%%%%%%%%%%

\subsection{The control system software}
\label{section_software}

While the CIRCUS heavily relies on the Sinara hardware to perform its operations, its core part is the software infrastructure. As introduced in section~\ref{section_intro}, it consists of two parts: ARTIQ and TALOS (both presented in greater details in the two following sections). ARTIQ is the high-level programming language for scripting the ns-precise routines to be executed by Kasli, that we empowered with libraries to streamline the coding of experimental routines and to integrate its operations with TALOS. In principle, the Sinara/ARTIQ structure could be integrated in different overall control system structures as well. In contrast, TALOS is the framework that constitute the slow-control: it both provides the interface between the operators and the apparatus, and its flexibility makes it compatible with any range of hardware and control software units independent of their precise characteristics.

It is in the interplay of this ns-precise hardware control on the one hand and its full integration and automation of the surrounding experiment on the other hand that the presented control system, CIRCUS, manifests its strength in such a way that it can be applied to any experiment with similar requirements.

This interplay is evident especially when it comes to executing a sequence of measurements. In fact, the schedule of scripts (with parameters) is defined using the apposite TALOS interface, and when the schedule is launched, it is TALOS that assesses if the conditions for running the experiment are met. If positive, it passes the command to Sinara, which executes the script, and TALOS remains available to forward calls from the used ARTIQ/Python script to any part of the experimental apparatus. When the script terminates, the command passes back to TALOS, which, based on the outcome of the script, decides what action is to be taken -- most of the time, running the same or the subsequent script in the schedule.

\begin{figure}[htbp]
\begin{center}
\includegraphics[width=0.9\textwidth]{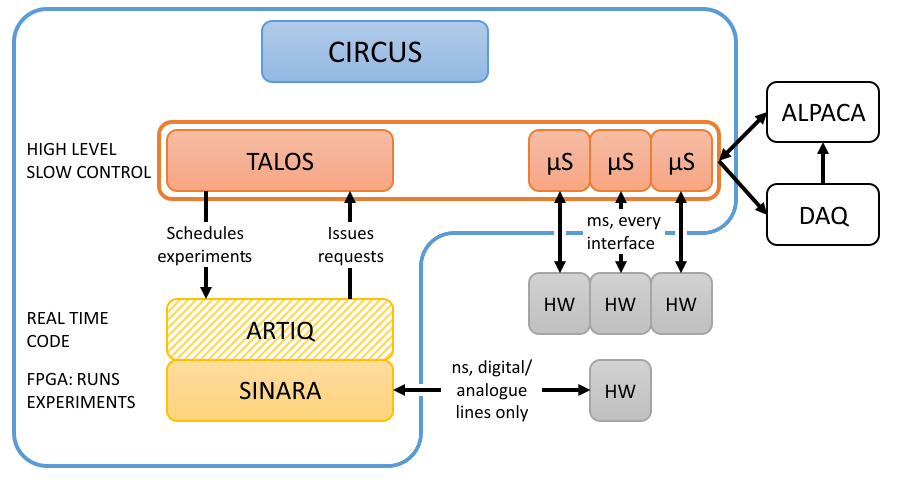}
\caption{Schematic of the CIRCUS control system and its constituent parts (ARTIQ/Sinara and TALOS), together with its relationship with other software and hardware subsystems.}
\label{fig_circus_scheme}
\end{center}
\end{figure}

In Fig.~\ref{fig_circus_scheme} the schematic of the CIRCUS control system is given, outlining the relationship of its constituent parts and their functionality, together with the connection with the other software and hardware components of \Aegis.

The CIRCUS control system is available open-source in a \emph{git} repository (DOI: 10.5281/zenodo.10371799)

In the following, both ARTIQ and TALOS are explained in greater detail.

\paragraph{ARTIQ} 
\label{section_artiq}

As explained in section~\ref{section_hardware}, the Sinara hardware relies upon the ARTIQ (Advanced Real-Time Infrastructure for Quantum physics) \cite{artiq:16} language for a straightforward, reliable software control. ARTIQ is a Python-based, high-level programming language which supplies specialised pre-generated functions for communicating with the hardware. The resulting control routines are formed by clear and short run scripts, preventing long familiarisation phases of semi-experienced programmers and allowing for quick adaptations during data taking.

ARTIQ is designed to script experiments with nanosecond resolution and microsecond latency. To meet the requirements of real-time programming, ARTIQ code consists of two parts which can interact with one another: the first one, composed of regular Python commands, is executed on the host, while the \emph{ARTIQ kernel} is executed on the CPU of the core device. This CPU can directly access a part of the ``gateware''\footnote{By gateware is meant the code specifying the configuration of the digital logic gates of an FPGA.} responsible for specialised programmable I/O timing logic. A timeline, constituted by all programmed input and output events, keeps the synchronisation of the experimental routines: output events with a given timestamp are executed in a first-in-first-out mode when matching an internal, high-resolution clock, and input events are recorded with a stamp for the current clock value.

The ARTIQ environment includes a dedicated function to observe a given I/O TTL channel and register rising or falling edge events for a specified duration. A sequence of actions can then be performed within a deterministically programmed time window after receiving a trigger, one example of this being another ARTIQ function designed to set a specified voltage on a given Fastino channel. In order to control multiple different trap electrodes in a synchronous way, the use of the provided Direct Memory Access (DMA) is essential, as it allows to pre-define RTIO sequences in the Kaslis' SDRAM, which can then be run directly by the FPGA core.

ARTIQ allows for a library-based approach to programming run routines of an experiment. To simplify and standardise the procedure for creating run scripts, an experiment parent class has been developed. All routines inherit from this main class, which contains both the code for initialisation and configuration of the hardware, and function libraries for interacting with the hardware and trigger signals, whose constituents can be called from the scripts defining the different experimental routines.
The effect of the outlined library-based approach can be observed in Fig.~\ref{artiq_code_example}, which shows a very simple experimental routine.  In both cases, the resulting sequence is the same: the system waits for an incoming trigger signal on one of the digital I/O lines and subsequently produces a voltage ramp to \SI{1}{\volt} on three of the Fastino channels (which is amplified to \SI{20}{\volt} by the corresponding amplifier units). The application of the calibration constants for the amplifier boards described in Appendix \ref{appendix_calibration} is included in the function used in the routine on the right. The functionality to set up and initialise the used hardware, which is part of the first two function definitions on the left, is included in the standardised \textit{Build} and \textit{Init} functions on the right. All other functions defined explicitly in the script on the left are included in the library structure and available without re-definition to all experiment scripts. This means that only one additional function call is needed in the actual run routine shown on the right side to achieve the same result as the code on the left.

The use of the \Aegis library system reduces the  ARTIQ script to a few lines of code when importing the parent classes and yields an immediate, simple overview of the routine. This effect is rendered more and more distinct the more complex (and closer to realistic run sequences) the experimental routines become.

\begin{figure}[htbp]
\hspace*{-1em}
\begin{minipage}[t]{0.4\textwidth}
\begin{center}
\begin{verbnobox}[\ssmall\hspace{1ex}]
    from artiq.experiment import *
    from artiq.coredevice.kasli_i2c import port_mapping

    class HVAamp_Trigger(EnvExperiment):

        def build(self):
            self.setattr_device("core")
            self.setattr_device("fastino0")
            self.setattr_device("ttl0")
            self.setattr_device("ttl_hvamp0_sw1")
            self.setattr_device("ttl_hvamp0_sw2")
            self.setattr_device("ttl_hvamp0_sw3")
            self.setattr_device("dio_mcx_dir_switch")
            self.setattr_device("i2c_switch0")
            self.dio_mmcx_i2c_port = port_mapping["EEM0"]

        @kernel
        def set_dio_outputs(self):
            self.i2c_switch0.set(self.dio_mmcx_i2c_port)
            self.dio_mcx_dir_switch.set(0b00000001)
            self.core.break_realtime()
            self.ttl0.input()
            self.core.break_realtime()

        @kernel
        def SignalAtTrigger(self):
            t_gate = self.ttl0.gate_rising(120*s)
            t_trig = self.ttl0.timestamp_mu(t_gate)
            at_mu(t_trig)
            delay(10*us)
            self.fastino0.update(1<<3|1<<2|1<<1)

        @kernel
        def SetVoltages(self):
            self.fastino0.set_dac(1, 1.0)
            self.core.break_realtime()
            self.fastino0.set_dac(2, 1.0)
            self.core.break_realtime()
            self.fastino0.set_dac(3, 1.0)
            self.core.break_realtime()
            self.SignalAtTrigger()

        @kernel
        def run(self):
            self.core.reset()
            self.fastino0.init()
            self.core.break_realtime()
            self.fastino0.set_hold(1<<3|1<<2|1<<1)
            self.SetVoltages()
        
\end{verbnobox}
\end{center}
\end{minipage}
\hspace*{-3em}
\hfill\vline\hfill
\hspace*{-5em}
\begin{minipage}[t]{0.4\textwidth}
\begin{verbnobox}[\ssmall\hspace{1ex}]
    import sys
    sys.path.insert(1, 'C:\kasli-code\Libraries')
    from AEgIS_imports import *
    from AegisExperiment import _AegisExpOfficial
    
    class HVAamp_Trigger(_AegisExpOfficial):
    
        def build(self):
            self.Build()
    
        def run(self):
            self.Init()
            self.SetVoltagesAtTrigger("Trigger", 10*us,
                "Channel1", 20.0, "Channel2", 20.0,
                "Channel3", 20.0)
\end{verbnobox}
\end{minipage}
\caption{Left: Experimental routine to set a specified output voltage on three amplifier channels of the Sinara hardware system after an incoming trigger pulse, programmed in the ARTIQ environment. Right: The same experimental routine as on the left, reduced to a few lines of code when implementing library-based programming.}
\label{artiq_code_example}
\end{figure}

In particular, a Python library, called the \emph{TCP Library}, has been created to organise the interface with the TALOS part of the control system infrastructure, containing the functions that ensure the communication between them. The TALOS system underwent an in-depth test during the antiproton run, exhibiting reliable interaction with the Sinara/ARTIQ setup. 

\begin{figure}[H]
\centering
\includegraphics[width=0.8\textwidth]{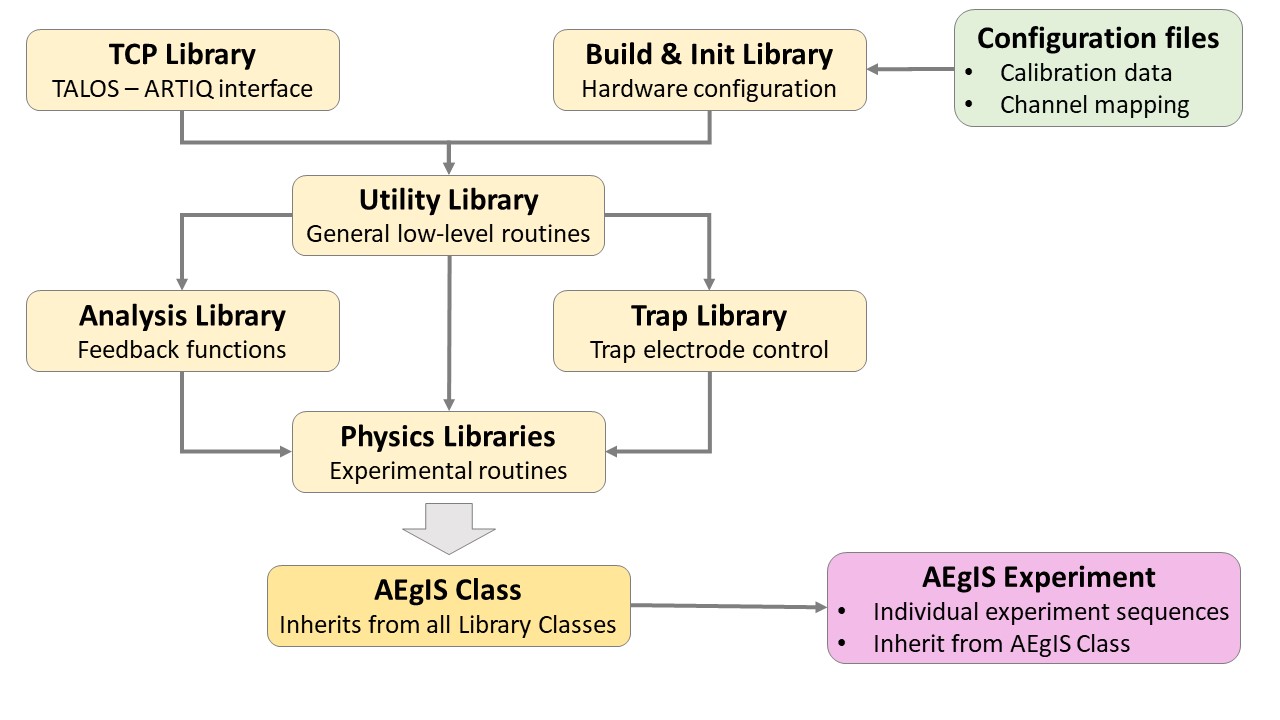}
\caption{Schematic of the ARTIQ/Python library structure of CIRCUS, as used in \Aegis. Each library defines a class, which all the experimental scripts of \Aegis inherit from. Most of the functions defined in the top-level libraries (\emph{TCP}, \emph{Build \& Init}, \emph{Utility} and \emph{Analysis} libraries) are generic and could be utilised by other experiments as well.}
\label{artiq_libraries}
\end{figure}

Fig. \ref{artiq_libraries} shows the library structure developed in ARTIQ/Python code that is used as the basis of the hardware communication of the presented control system. Each shown library is formed by a class, which the AEgIS Class, i.e. the parent class of the experimental scripts, inherits from. As shown in the schematic, the higher-level libraries use functions of the base classes. The actual run routines are then sub-classes of the AEgIS Class and have all library methods available. Of course, several of the functions, particularly in the lower, experimental libraries, are specific to the \Aegis experiment and would need to be replaced by corresponding functionalities in other environments. On the other hand, the base functions in the \emph{TCP Library}, used to interface with TALOS, as well as the standard routines to configure and initialise the used hardware (with adapted configurations) and those general functions related to timing synchronisation, information logging, and data retrieval in the \emph{Utility Library} and \emph{Analysis Library} are re-usable as general functionalities of CIRCUS.

\paragraph{TALOS}
\label{section_talos}

TALOS (Total Automation of LabVIEW\texttrademark\ Operations for Science) is a control system framework\footnote{We refer to TALOS as a \emph{framework} because it does not only come with the functionalities described in this section, but it also creates a specific way of coding, in the form of guidelines to write µServices.} that unifies all the computers of an experiment into a coherent, coordinated, distributed system, and it increases the reliability and stability of the running apparatus by means of a (distributed) watchdog structure, with the ultimate goal to safely leave it running unsupervised for extended periods of time. 

It is founded on two concepts: the ``everything is a µService\footnote{Read \emph{MicroService}.}'' approach, and the distributed architecture. To satisfy both requirements, it was decided to base TALOS on the Actor Model~\cite{hewitt:73}, which is an information theory model specifically designed for the implementation of large distributed system architectures. The theory is based on the concept of \textit{actors}, single entities that can react to a message arriving from another actor by executing a local action, sending further messages to other actors, changing its internal status, creating additional actors, or a combination of the above. 

The first concept, ``everything is a µService'', consists of the division of the code into independent, autonomous parts, the µServices, each with a defined scope and task. Each µService runs separately from the other µServices, in a completely asynchronous way, communicating among each other via a built-in messaging system. This design choice makes the system both easily extendable and debuggable in a straightforward way, while also minimising system downtime: in fact, every µService can be separately tested before being deployed, and any problem can be readily isolated and solved.

The second concept, the distributed architecture, manifests in multiple instances of the same actor, called \textit{Guardian}, taking the role of root actor, one on every computer. This Guardian has the function of monitoring both the status of all µServices running locally, all implemented as independent actors, and the status of the other active Guardians in the network. At the same time, the Guardians provide a common infrastructure to share messages and data between various µServices and among different computers. This new paradigm has a twofold result: it strengthens the reliability, the safety, and the stability of the system through a distributed watchdog system (in fact, no computer or program can become unresponsive without it being noticed), and it unifies all the computers into a single, distributed entity. The latter is what facilitates the full automation of the experimental procedures, as high-level decisions often depend on parameters generated by multiple computers.

The choice to base this new framework on LabVIEW\texttrademark\ (by NI\footnote{Formerly \textit{National Instruments}.}) was dictated mainly from the fact that an implementation of the Actor Model is present in LabVIEW\texttrademark, called \emph{NI Actor Framework}, which provides a readily available foundation block. Moreover, in \Aegis, as in many other experiments, some fundamental hardware components are from NI, and therefore natively interfaced in LabVIEW\texttrademark, simplifying µServices coding.

Some µServices developed with TALOS are of general use, independent of the \Aegis experiment: CIRCUS comes with them integrated, so to be readily utilised by other experiments. Aside the µServices managing the communications with the FPGA (more below) and parts of TALOS internal mechanics itself, some good examples are: the \emph{Error Manager}, which serves as a the single concentrated point for all the errors of the distributed system; the \emph{Scheduler}, used as an interface for the user to define sequences of experimental scripts, each with specific parameters; the \emph{Monkey}, which executes the scripts in the schedule and takes the high-level decisions at the core of the automation of the control system (such as retrying a script if it did not run correctly, or modifying the parameters based on the feedback from the analysis system); and the \emph{Tamer}, used to coordinate the parallel execution of multiple \emph{Monkeys}, in case multiple Sinara/Kasli crates need to be managed simultaneously.

As stated before, for the seamless functioning of the CIRCUS control system, a critical part of TALOS is the interface with Kasli. In fact, naturally, Kasli is managed by a user via a command-line interface from a terminal, and communication with external hardware is foreseen to happen only via its digital lines. In more complex experiments like \Aegis, though, Kasli needs to communicate a huge variety of messages towards multiple different systems, in order to keep all the hardware operations synchronous, and this is impossible to be realised via physical digital lines, the more so because the messages often carry a non-trivial data structure. TALOS, in this respect, provides an interface to the FPGA to extend its capabilities: thanks to a direct TCP (Transmission Control Protocol)~\cite{tcp} communication between Kasli and two dedicated µServices, the FPGA can send (and receive) string messages to (and from) all µServices. This enables Kasli to have full ``slow''\footnote{The messages run over the network, so the speed of communication is inherently on the order of the \SI{}{\ms}.} control over all the hardware and software interfaced with TALOS, which would be impossible by leveraging only Kasli native capabilities.

In addition, the usual terminal communication with Kasli is also integrated with TALOS via a specific µService, called \emph{Kasli Wrapper}. It provides a low-level interface to communicate with it in a native manner, useful in case the TCP connection is not available (before the instantiation of the latter, or in case of errors).

This solution, coupled with a few digital lines controlled by the FPGA, enables the correct synchronisation of complex operations (e.g. setting the potential of a specific electrode to a specific voltage, configuring and starting the acquistion of a spectrometer) with a precision in the order of \SI{}{\ns}.

As mentioned before, TALOS could be easily modified in order to integrate a different real-time system. In fact, the terminal communication with a different FPGA can be simply assimilated by creating a child of the \emph{Kasli Wrapper} µService, and coding it to redirect the messages between TALOS and the new FPGA. Similarly, to leverage the power of the TCP connection, the base functions present in the ARTIQ \emph{TCP Library} have to be reimplemented in the language used by the new real-time system, or readily used if Python is supported.

The TALOS framework itself is the subject of a dedicated publication~\cite{talos_2024}.

Moreover, also TALOS is freely available open-source in a \emph{git} repository (DOI:~10.5281/zenodo.10371404).

\begin{sidewaysfigure}
\centering
\includegraphics[width=0.99\textwidth]{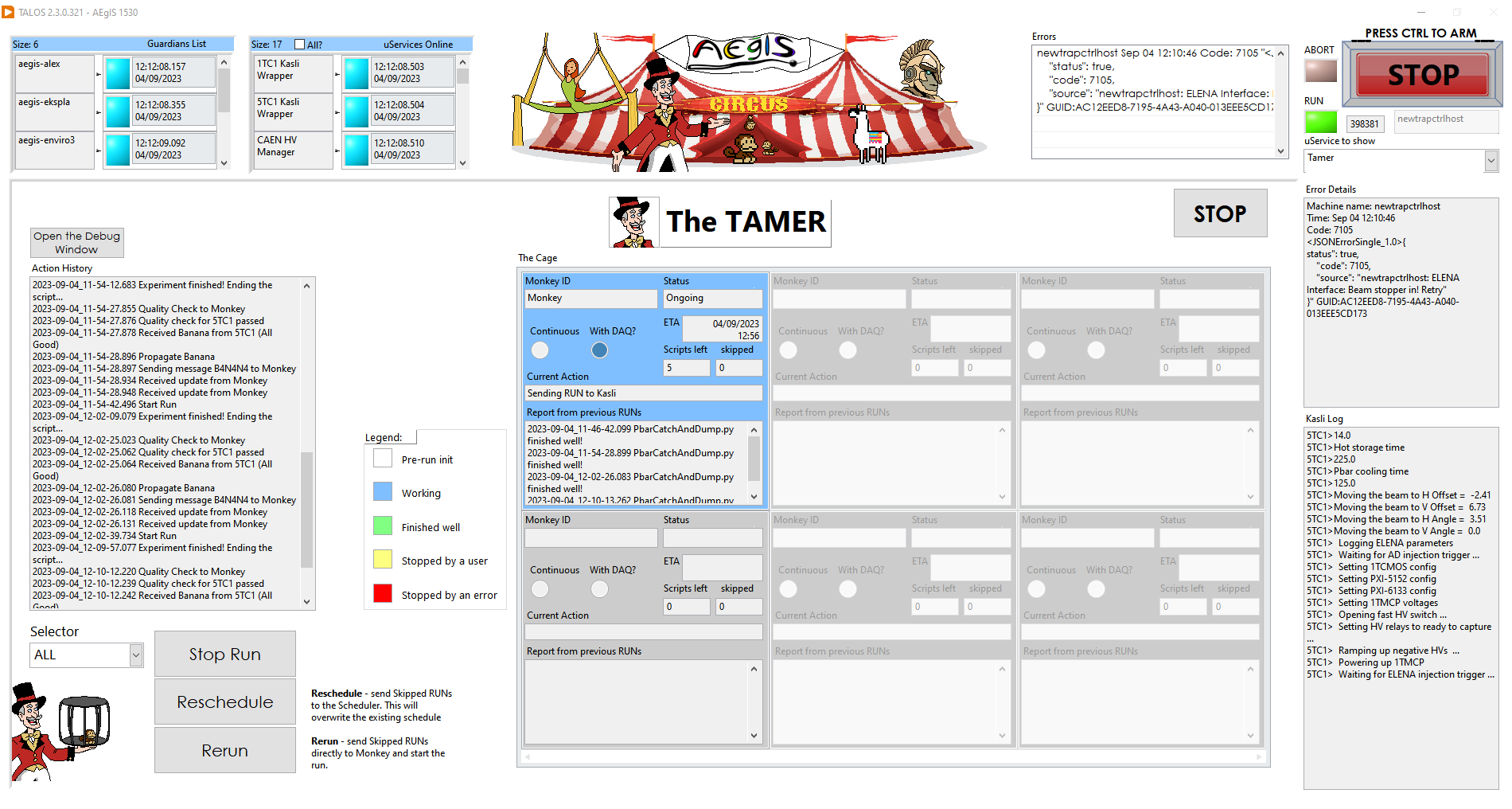}
\caption{A view of the CIRCUS control system running a schedule of experiments with antiprotons. This main window is provided by TALOS. On the top right the \emph{Guardians} and $\mu$Services watchdogs are visible, while on the top left the error list is present. On the right column there are, on top, the details of the error selected, and below, the live log of the operations of Kasli. This frame is common and identical on all the machines of the experiment. In the main window the Tamer $\mu$Service is displayed, which is currently monitoring one schedule of measurements running.}
\label{TALOS_screenshot}
\end{sidewaysfigure}

%%%%%%%%%%%%%%%%%%%%%%%%%%%%%%%%%%%%%%%%%%%%%%%%%%%%%%%%

% Data Acquisition

%%%%%%%%%%%%%%%%%%%%%%%%%%%%%%%%%%%%%%%%%%%%%%%%%%%%%%%%

\subsection{Data acquisition}
\label{section_DAQ}

\begin{figure}[htbp]
\centering
\includegraphics[width=0.8\textwidth]{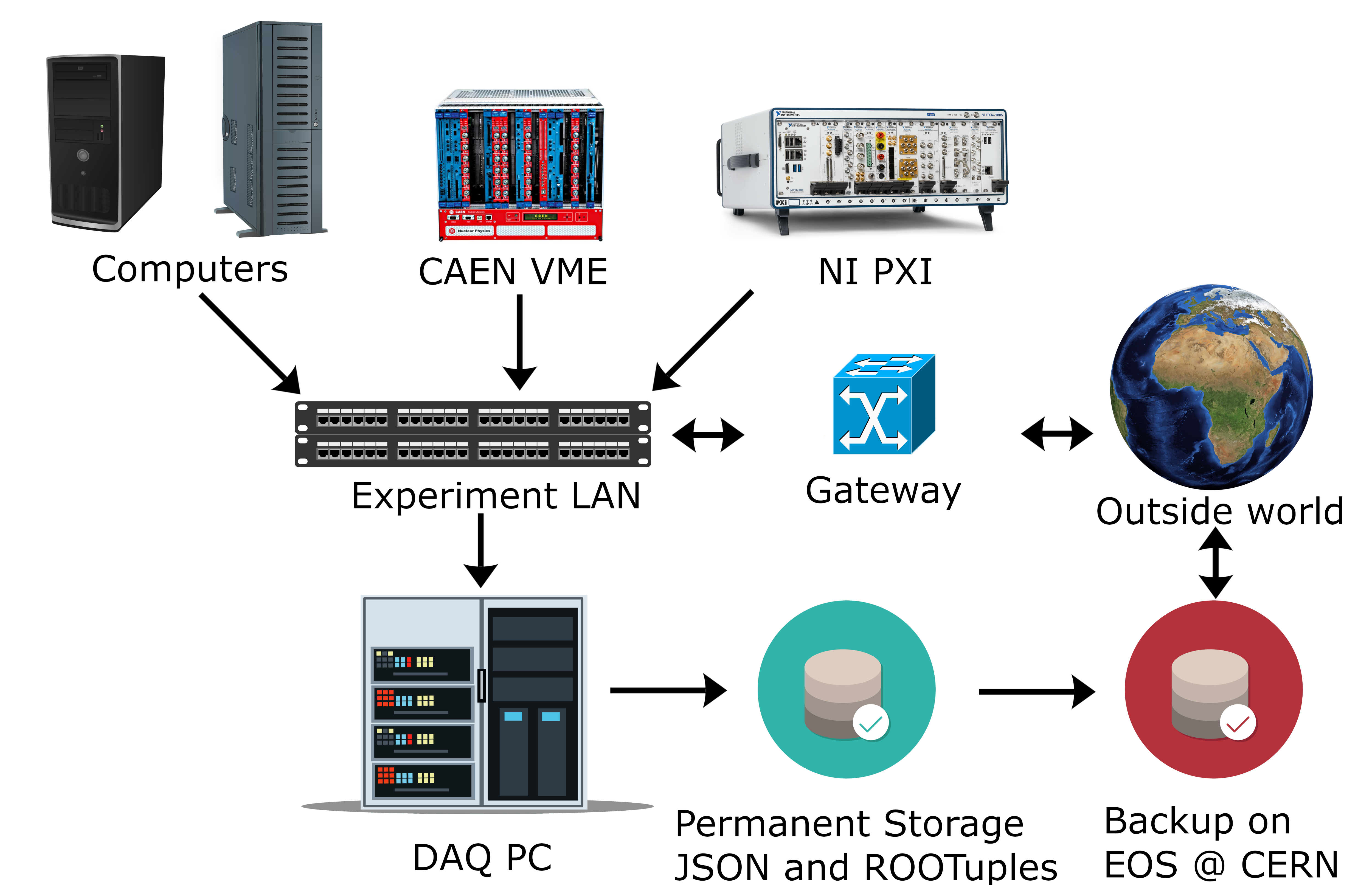}
\caption{Schematic of the data flow in \Aegis. All devices (computers, VME and real-time) are connected to a common LAN subnet and send data to the DAQ PC as GXML Data Objects over TCP or SCP (Secure Copy Protocol). The DAQ computer permanently stores the data on hard-drives as JSON files and ROOTuples. A further backup copy of the data is generated on EOS~\cite{EOS_Peters_2011} at CERN. The data can be accessed from outside CERN from EOS or directly from the DAQ computer via a dedicated gateway.}
\label{fig:daq-interaction}
\end{figure}

Every experiment has the need to save and store the data collected during the measurements. For this purpose, the \Aegis experiment operates an integrated run and monitoring data acquisition (DAQ) and logging system. Data atoms\footnote{The term \emph{data atom} refers to one unit of the smallest data container used.}, all cast in the standard format described in Table~\ref{tab:datom}, are generated at various locations in the experiment, transferred over the local-area network (LAN), saved to local storage, then saved to long-term disk and tape storage systems at CERN.
Data sources and sinks, along with the data transfer paths over the LAN, are identified in Fig.~\ref{fig:daq-interaction}.
This system is designed for the vital parameter monitoring needs for experiment commissioning and the long-term data logging for experimental runs, and has been running for over a decade.

\begin{table}[htbp]
\centering
\begin{tabular}{|c|l|}\hline
{\bf Name} & {\small Alphanumeric String containing a (possibly hierarchical) unique}\cr
           & {\small name for the data atom. The format of the data associated}\cr
           & {\small with a given name should not be changed.}\cr \hline
{\bf Timestamp} & {\small Instant when the data was acquired, in three formats:}\cr
           & {\small 1) character string, parsable by {\tt strptime(3)};}\cr
           & {\small 2) {\tt struct timespec} containing time since the UNIX epoch;}\cr
           & {\small 3) 64-bit unsigned integer with RF clock count, if applicable.}\cr\hline
{\bf Data} & {\small Instance of a scalar, vector, or structured (cluster) data type,}\cr
           & {\small compatible with LabVIEW\texttrademark\ types, and their conversion}\cr
           & {\small into either JSON-formatted files or ROOT {\tt TTree}s.}\cr\hline
\end{tabular}
\caption{Structure of the \Aegis data atom, representing {\it all} DAQ data objects.}
\label{tab:datom}
\end{table}

The data are saved in JSON-formatted files, which provide a compact, clearly structured standard for efficient generation and transfer and are compatible with the GXML reference library (for serialisation) of the LabVIEW\texttrademark\ architecture used in many experiments.

For online access of monitoring data, CERN's ROOT data format system is currently still used preferentially thanks to its high data compression functionalities.

A side-by-side comparison of text representations of the general-purpose \Aegis data atom in the GXML and JSON formats is shown in Fig.~\ref{fig:gxmljson}.

\begin{figure}[htbp]
\begin{minipage}[t]{0.5\textwidth}
\begin{verbnobox}[\ssmall\hspace{1ex}]
    <GXML_Root>
        <Name type='String'>
            test_cluster
        </Name>
        <Timestamp mems='4'>
            <str type='String'>
                16:18:09.220036 09/20/2016
            </str>
            <tv_sec type='U64'>1474381089
            </tv_sec>
            <tv_nsec type='U32'>220036174
            </tv_nsec>
            <Clock type='U64'>7856432</Clock>
        </Timestamp>
        <Data mems='3'>
            <double_val type='DBL'>
                1.2344999999999999307
            </double_val>
            <int_val type='I32'>12345</int_val>
            <float_array dim='[3]' type='SGL'>
                <v>1.1</v><v>2.2</v><v>3.3</v>
            </float_array>
        </Data>
    </GXML_Root>
\end{verbnobox}
\end{minipage}%
\hspace*{-5em}
\hfill\vline\hfill
\hspace*{-9em}
\begin{minipage}[t]{0.5\textwidth}
\begin{verbnobox}[\ssmall\hspace{1ex}]
    [ { "test_cluster": {
        "Timestamp": {
            "clock": 7856432,
            "str":
                {\scriptsize "16:17:18.020212 10/20/2021"},
            "tv_nsec": 20212223,
            "tv_sec": 1634739438
        },
        "Type": "",
        "double_val": {
            "Type": "DBL",
            "__value": 1.2345
        },
        "float_array": {
            "MemberDims": "[3]",
            "Type": "Array",
            "v": [
                1.100000023841858,
                2.200000047683716,
                3.299999952316284
            ]
        },
        "int_val": {
            "Type": "I32",
            "__value": 12345
    } } } ]
\end{verbnobox}
\end{minipage}%
\caption{Left: Example of GXML serialisation of an \Aegis data atom containing
a cluster of two numeric scalar values and one numeric array. Right: The corresponding JSON equivalent representation.}
\label{fig:gxmljson}
\end{figure}

The presented DAQ system was built and adapted according to the specific needs of the \Aegis experiment and is explained here for completeness. Other data acquisition systems, based on different hardware and software setups, can of course be easily integrated in the overall control system structure analogously. Provided that the data acquisition system supports an interface with the commands \emph{Start}, \emph{Stop} and \emph{Send data}, its integration in CIRCUS would simply consist of creating a child of the \emph{DAQ Manager} µService, and implementing inside it the proper interface with these commands. After that, TALOS and all the other µServices will immediately use the new data acquisition system for data saving, without any further change in the code.

%%%%%%%%%%%%%%%%%%%%%%%%%%%%%%%%%%%%%%%%%%%%%%%%%%%%%%%

% ALPACA

%%%%%%%%%%%%%%%%%%%%%%%%%%%%%%%%%%%%%%%%%%%%%%%%%%%%%%%

\subsection{Integrated analysis pipelines} 
\label{section_ALPACA}

Analogously to the data acquisition system, every experiment also has the need for a series of algorithms to analyse the obtained data. Often, part of the data analysis is used to tune and improve the subsequent data acquisition: the capability of a control system to perform this task in autonomy is of great advantage to the scientists.

\begin{figure}[htbp]
\begin{center}
\includegraphics[width = 0.9\textwidth]{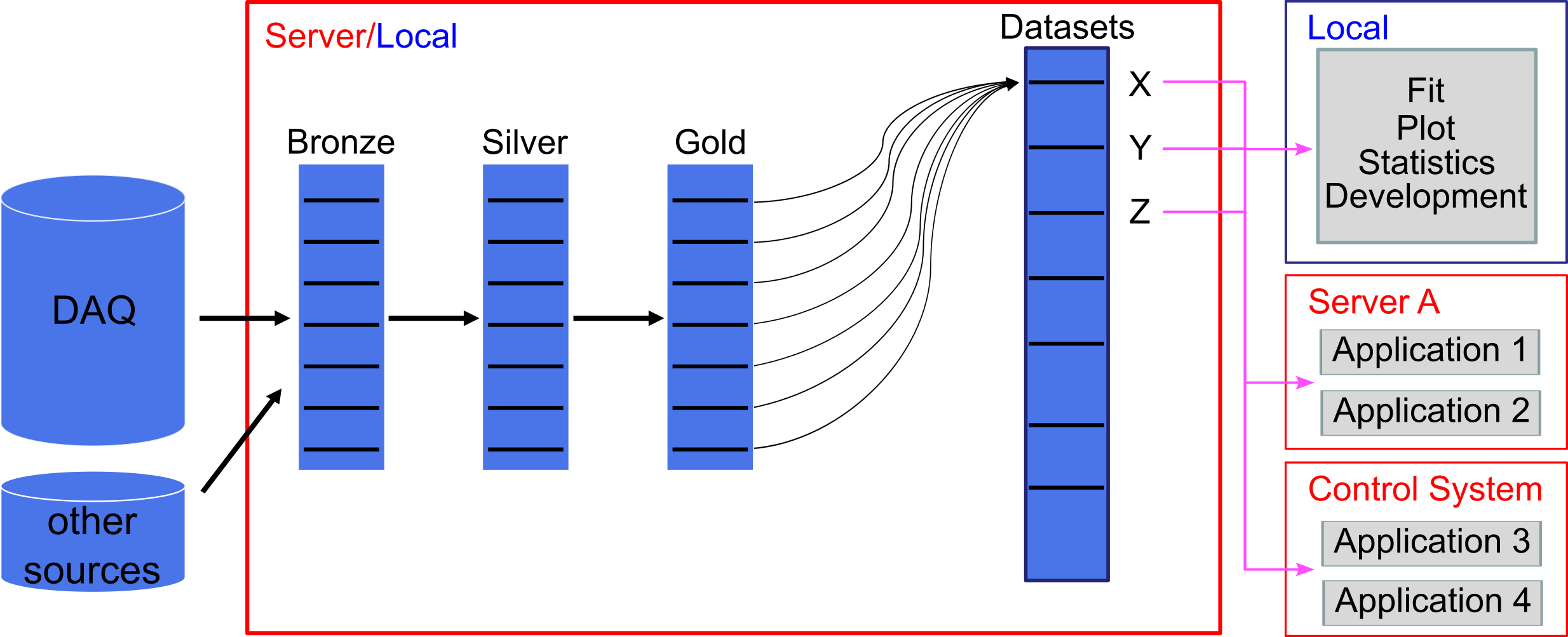} 
\caption{Representation of the architecture of the ALPACA analysis framework, including the stepwise processing of the data as well as the local or server based deployment.}
\label{python_architecture}
\end{center}
\end{figure}

All Python Analysis Code of \Aegis (ALPACA) is a Python data analysis framework written specifically for the \Aegis experiment’s different physics tracks. It leverages the functionality of the \textit{NumPy}~\cite{numpy}, \textit{SciPy}~\cite{scipy} and \textit{Plotly}~\cite{plotly} libraries to collect and transform the raw data acquired by all the detectors into observables, which can then be utilised by the scientists to perform dedicated studies. Figure~\ref{python_architecture} depicts the framework’s linear architecture, where pipelines transform the data into different processing states. 

First, all raw sources of an experiment’s data, stored on different servers and in different formats (e.g. ROOT, json, png, txt, etc., and either originally plain or zipped) are concatenated into a bronze state as a Python dictionary. Raw sources include the data of each detector triggered, the settings of the detectors and the environmental data (for example, temperature and vacuum readings) during the experiment. At this stage, the originally stored files are just saved in a Python native format but no data manipulation is applied.

From the bronze to the silver state, the data is restructured depending on how each detector stores the acquired data according to its own configurations. For example, the json files for the acquired voltage readout of the MCP detector\footnote{A Micro-Channel Plate used to detect particles at the far end of the \Aegis experiment. The electrons generated are converted into light by a phosphor screen and imaged with a camera. The voltage profile of the MCP itself is also acquired.} always contain, as the first and second entries, the start time of the acquisition and the time increment, while the remaining entries hold the actual voltage readings after each time increment. In the bronze~$\rightarrow$~silver pipeline, these data are parsed such that the start time, the time increment and the voltage readout become variables accessible on their own. Moreover, a three-layer nested data structure is established with the detector on the top, the acquisition number in the middle and the acquired data and run-specific configurations of the detectors at the deepest level. 

Subsequently, the silver~$\rightarrow$~gold pipeline computes and appends observables for each detector and acquisition, still preserving all the original data is preserved in the gold state as well.
For example, in this step, the image taken from the MCP camera is first normalised for the set gain of the MCP, then the background is evaluated and subtracted, before sum, mean and standard deviation of all pixels are calculated and made available as three different observables.

In the last step, user-specified datasets of observables over many experiments are concatenated and made available for the user’s personal analysis as well as for applications. Additionally, a dedicated package for the generation of statistics fits and plots as well as for the training, evaluation, and use of machine learning models using the generated datasets has been developed.

Thanks to the single end-point for querying datasets from ALPACA as well as the independence of the pipelines from each other, ALPACA is easily scalable in the number of applications as well as in the data sources and processing pipelines. Special emphasis is put on the scalability and reusability of the source code, which allows the seamless integration of new detectors installed at the \Aegis apparatus as well as new analysis pipelines.
Different applications utilising ALPACA’s end-point for datasets beyond simple user’s analyses have been envisioned for the future, enabling especially the introduction of automated feedback loops via the main control system to autonomously take decisions and promptly adjust the experimental settings for the subsequent experiment. Such feedback loops can be used for optimisation problems and event triggering, thereby increasing the overall progress speed of \Aegis by integrating the ALPACA framework directly into CIRCUS as well.

Table \ref{analyses_runtime} includes samples of the current runtime performance on a set of 177 experiments, which produced an average of 21.4 MB of raw data.

\begin{table}[!h]
\begin{tabular}{ccccc}
\hline \hline 
\multirow{2}{*}{Number of experiments} & \multicolumn{4}{c}{Loading times, in seconds, from:}               \\ 
                                & Raw     & Bronze     & Gold     & Datasets    \\ \hline 
1                               & 3.96    & 1.42       & 0.22     & 0.009        \\
10                              & 36.20   & 15.86      & 2.33     & 0.015        \\
100                             & 380.27  & 176.76     & 24.47    & 0.06         \\
177                             & 658.81  & 310.85     & 46.04    & 0.09         \\ \hline \hline

\end{tabular}
\caption{Runtime performance of the analyses framework using the experimental data from a parameter scan during the antiproton beam time 2022. These times are characteristics of the \Aegis system.}
\label{analyses_runtime}
\end{table}

A significant speed-up in development and analyses is achieved by reloading the data from the different processing states. Loading the data from “Raw” takes exceptionally long due to the necessary download from the \Aegis servers, while the locally stored datasets are available almost instantaneously. Processing the data of a single experiment usually takes few seconds, which is feasible for feedback loops with the control system.

In the framework of the presented control system, ALPACA is a powerful tool to aid automation and enable self-optimisation, and it is used as the main analysis framework in \Aegis. In principle, its design serves as a foundation and its use can be adapted to different experiments as needed. However, different software architectures that fulfil this purpose can also be used in its place. In particular, the capability of CIRCUS to autonomously modify the experiment parameters based on the feedback loop from the data taken (an example of which is given in the following section) relies on a simple interface with the analysis framework. It consists of two shell commands: one for retrieving the last measured value of a specified observable, and another one for receiving new parameters to use, given a list of parameters used and results obtained. Any analysis framework capable of producing such a simple interface would be straightforwardly integrable in CIRCUS.

%%%%%%%%%%%%%%%%%%%%%%%%%%%%%%%%%%%%%%%%%%%%%%%%%%%%%%

% Automation of lasers

%%%%%%%%%%%%%%%%%%%%%%%%%%%%%%%%%%%%%%%%%%%%%%%%%%%%%%

\subsection{First automation with feedback loop: timing stabilisation of a laser pulse}
\label{section_automation_laser}

The combination of the new control system and the new framework for data taking, storing, and pre-processing yields another desirable feature: decision-making based on a feedback loop. Complex systems typically depend on a multitude of parameters, of which not all are directly controllable. 

A good study case is the stabilisation of the pulse timing of one of the \Aegis{} lasers. In fact, the \Aegis laser system for positronium excitation to the n$\,=\,2$ state displays a strong correlation between ambient humidity and the resulting generation instant of the light pulse. The humidity in the environment, on the other hand, is coupled to the temperature, which in turn affects the output laser energy. Since the current ``climate control'' system can either stabilise the humidity or the temperature, the other needs to be allowed to run freely. 
The nanosecond-precise control system opens up the opportunity to tune the timing of the laser pulse by means of triggering a Pockels cell at the right moment, whereas the energy of the laser cannot be adjusted that easily. Thus, the temperature (and consequentially the energy of the laser pulse) is chosen to be controlled by the climate system, while the humidity is left to run freely. 
In turn, the time drift caused by the humidity variation is compensated by the control system via a feedback loop, which is detailed below.

\begin{figure}[htpb]
\begin{center}
\includegraphics[width = 0.75\textwidth]{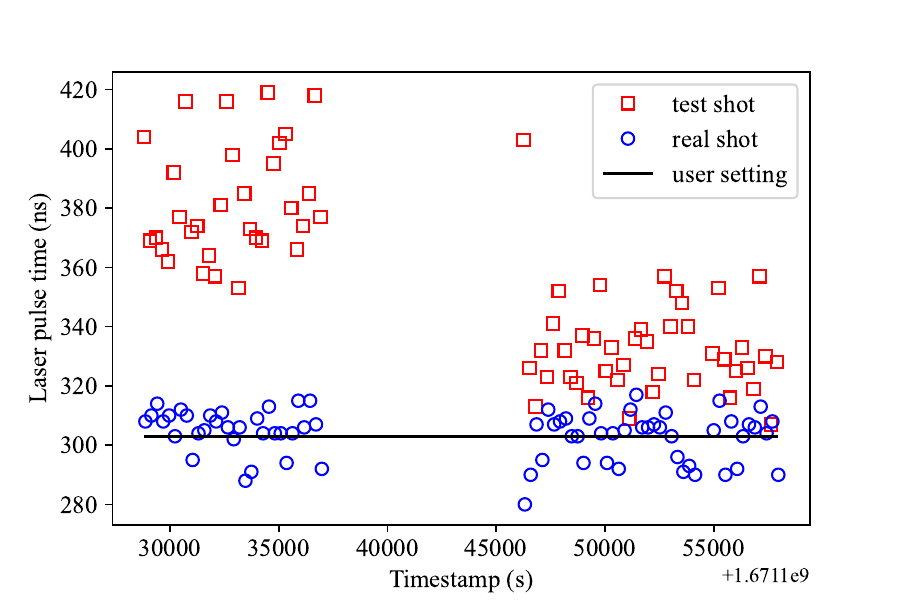} 
\caption{A feedback loop uses the uncorrected laser pulse timings (red squares) to calculate the deviation from the user setting (solid black line) over the course of an hour, and corrects the timing of the subsequent desired laser pulse that is used for the actual experiment (blue circles). Independent of short-term to long-term drifts or even sudden jumps, the resulting timing is always close to the desired value.}
\label{Alex_jitter_corr}
\end{center}
\end{figure}

A few seconds before the actual positronium production instant, a test laser pulse is produced by triggering the Pockels cell and the data acquisition chain. The generation instant of this pulse, depending on the environmental conditions, may vary with respect to the moment the Pockels cell is triggered, e.g. because humidity drifts over time. The acquired spectrum of a photo diode is immediately stored by the DAQ system and analysed by a dedicated function in the experimental script, which extracts the arrival time of the test laser pulse. It is then compared to a user-defined value and a correction term is calculated. Imminent to positron implantation into the converter target, the Pockels cell is triggered again for the actually used pulse, applying the correction term obtained from the test pulse to account for the temporal offset. As a result, the synchronisation is now sufficiently precise to guarantee an overlap of the laser pulse and the positronium cloud, independently of the origin of the drift. This can be seen in Fig.~\ref{Alex_jitter_corr}, where the timings of the test laser pulses (red squares) and the desired laser pulses (blue circles) are plotted for a series of experimental trials executed over the course of one hour (with some interruptions). The user-defined value is given as the horizontal line. The statistical errors on the determination of the timings are of the order of a few hundred picoseconds and thus not visible in the plot.

This active feedback loop, exemplified for the timing of a laser pulse, is versatile and can be applied to any parameter of any part of the system, given that there is enough time to obtain the test data and analyse it before the real experiment occurs. With this step, the control system becomes self-governed and self-stabilising, obtaining the ability to tune parameters autonomously for an optimal result. 

%%%%%%%%%%%%%%%%%%%%%%%%%%%%%%%%%%%%%%%%%%%%%%%%%%%

% Results

%%%%%%%%%%%%%%%%%%%%%%%%%%%%%%%%%%%%%%%%%%%%%%%%%%%

\section{Results and Discussion}
\label{section_results} 

Throughout the data taking of the 2021 antiproton beam time, three computers and two Sinara crates were used to perform the experiments. The computers were executing the CIRCUS control system and running in total 17 µServices, and they operated continuously during the whole period of beam time. 
Although the system was de facto undergoing its first field test campaign, it exhibited a very good stability, with an up-time close to 100~\% of all the foreseen beam time. Moreover, albeit not yet complete, the new control system already proved capable of operating the \Aegis experimental apparatus and routines in a completely unsupervised mode: in fact, it ran in an unmanned way throughout all nights of the data taking. In addition, the automation was advanced further to perform parametric scans within multi-dimensional phase-space: again, the system displayed the ability of running up to (in total) $\sim$~1000 data points over four different parameters, autonomously pausing and resuming the measurements when detecting manageable exceptions, e.g. when there was an interruption in the beam delivery from ELENA.

In 2022, the control system was further upgraded and refined, rendering it more stable, with a better error management and handling of external events (i.e. retrying a run if a µServices could not contact the DAQ, or if the beam of antiprotons from ELENA was empty). A total of six PCs were running more than 100 µServices (some of them were multiple instances of the unique 42 µServices coded). Apart from down time due to technical development on the experiment (or the decelerator complex), the system took data continuously. The interface with the Sinara electronics has been refined to allow for the option of using multiple, independently running units simultaneously. This feature will become critical once antihydrogen is routinely produced in large numbers. Ultimately, the integration of the analysis framework has enabled the system to autonomously derive certain values of some parameters of the experiment, based on a feedback-loop-driven machine-learning optimiser. This has completely changed the operation modality: from long scans and offline analyses to find the best working settings, to programming the machine to actively, and continuously, find them in an autonomous way.  

The triggers from AD/ELENA were reliably registered by the digital I/O units of the Sinara crates and propagated through the control system to all involved hardware. The working principle was the following: upon reception of the “ELENA Injection” trigger (which arrives approximatively \SI{30}{\second} before the antiprotons actually reach the experiment) by Kasli, all the hardware systems are initialized and prepared to respond to a trigger signal, which is then given from a digital line of Sinara upon reception of either the “Bunch arriving – \SI{20}{\us}” (for slower hardware such as cameras) or the “Bunch arriving”\footnote{The time difference between the arrival of this signal and the effective arrival of the particles is settable by the experiment, and it is typically in the order of hundreds of ns.} (for fast hardware such as high voltage electrode gates) signals.

\begin{figure}[htbp]
\centering
\includegraphics[width=0.9\textwidth]{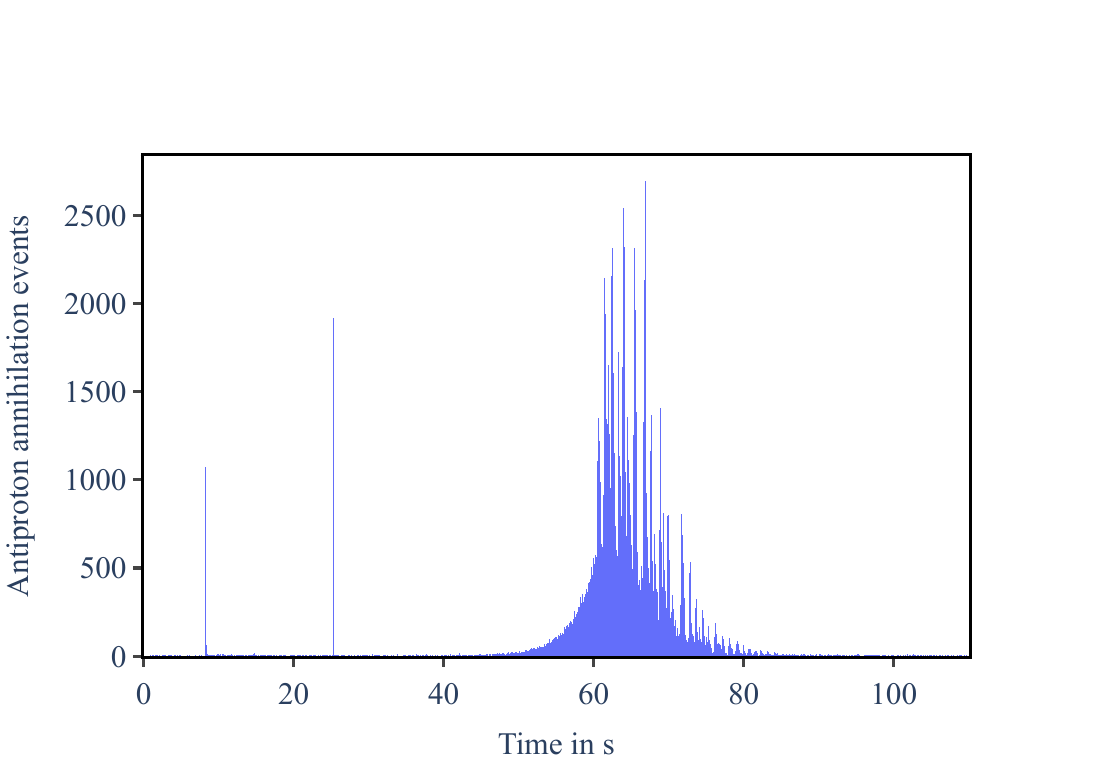}
\caption{Scintillator counts of the annihilation of antiprotons inside the Penning trap. The two deltoid-like structures at \SI{\sim10}{\s} and \SI{\sim25}{\s} are emissions from the accelerators in the AD complex and are present independently of the ongoing trapping.}
\label{pbar_trapped}
\end{figure}

Thanks to the features of the CIRCUS control system as well as other recent improvements to the experiment, antiproton capture in the trap was efficiently performed: the synchronisation capabilities provided by Sinara, coupled with the fast iteration regime facilitated by ARTIQ, enabled a fully parameter-optimised capture of the energy-degraded portion of the antiprotons in less than 10 days after the first beam was acquired\footnote{For comparison, a similar optimised results was achieved with the previous system in more than 3 months.}. To monitor the capture efficiency, three different scintillating fibres, each connected to a photomultiplier tube (PMT), were used: by operating the PMT in the non-saturation regime, the quantity of antiprotons was estimated by the amplitude of the detectors’ signals. The difference in the signals between measurements without raising the electrode gates (``passthrough mode'') and with the electrodes raised at the correct time (``capture mode'') confirms the capture of a significant amount of antiprotons available from ELENA (preliminary estimates point towards a record trapping efficiency around 70~\%~\cite{aegis_spsc:23}). As shown in Fig.~\ref{pbar_trapped}, the annihilation signals in the surrounding scintillators indicate trapping of antiprotons for up to \SI{50}{\s}, a lifetime in agreement with the initially very poor vacuum level ($\approx 10^{-8}$~\SI{}{\milli\bar} at the time of this measurement) and the absence of electron cooling. The characteristic bell shape of the annihilation events is given by the fact that initially, the antiprotons are trapped at several \SI{}{\kV} energy, but the cross-section of annihilation is effectively greater than zero only for energies in the order of tenths of \SI{}{\eV}. Therefore, at the beginning, no annihilations take place because the antiprotons are losing energy by elastic collision with the residual gas. When their energy is low enough, they start to annihilate, which here happens from around \SI{45}{\second}. Since the population of low-energy antiprotons increases in time, the annihilation count rises, reaching a peak when the depletion of the antiprotons in the trap starts to be significant. From there, the curve decreases, terminating with an exponential decay with the characteristic lifetime of the cold antiprotons in the trap.

In parallel, and unrelated to the experiments performed with antiprotons, the experiment employs two laser systems, the so-called ``EKSPLA'' (\SI{205}{\nm} and \SI{1064}{\nm}), which is a Nd:YAG pump-based system for antihydrogen formation, and an alexandrite-based system (in the following referred to as ``Alex'', \SI{243}{\nm}), used in experiments with positronium. These two setups are operated independently from each other, and they are spatially separated by more than \SI{5}{\m}, but during measurements, it is essential to keep them synchronised. This has been achieved by taking advantage of the master-satellite operation mode of Kasli devices. Two configurations have been tested: the continuous, standalone operation, and the on-demand operation. In the first scenario, the EKSPLA pulses at a frequency of \SI{10}{\Hz} are synchronised with the \SI{4}{\Hz} Alex pulses by a couple of Sinara crates, kept in master-satellite configuration through an optic fibre connection. On the master, an idle script continuously runs without the need of a computer and simultaneously re-triggers both lasers every \SI{30}{\s}, so as to temporally realign them and to eliminate any accumulated drift. In on-demand operation mode, by contrast, the lasers and the Sinara crates are kept idle, and the user can, at will, run a script which synchronously activates the pumping of the two lasers and subsequent simultaneous triggering.

%%%%%%%%%%%%%%%%%%%%%%%%%%%%%%%%%%%%%%%%%%%%%%%%%

% Conclusions

%%%%%%%%%%%%%%%%%%%%%%%%%%%%%%%%%%%%%%%%%%%%%%%%%

\section{Conclusions} 

The \Aegis collaboration has implemented CIRCUS, a novel, high level and very general system for controlling complex physics experiments based on the Sinara/ARTIQ open hardware/software ecosystem and the TALOS software infrastructure.

The first in-depth stress tests of the new control system during the regular antiproton run time at CERN's Antiproton Decelerator have successfully validated its usability for ns-precise synchronisation of the involved procedures and its continuous reliable operation. It demonstrates sound reproducibility of the experiments.

Consequently, the control system will be extended further to include additional parts of the experiment and enable an autonomous execution of the more complex activities foreseen for future beam times. Additionally, the interface between TALOS and ARTIQ will be improved with a more advanced library structure and the possibility of operating multiple Sinara units simultaneously, but with different modes of synchronisation among them (i.e. some running synchronously, and others asynchronously). A higher level of online data analysis integration is being implemented, since the new optimisation-driven approach is significantly improving the operation modality, reducing the beam time required and enabling manoeuvres previously unfeasible.

By providing such automation of the entire run operation, the CIRCUS control system will continue to optimise the uptime and quality of data taken during the upcoming measurement campaigns of \Aegis, including complex experiments such as the formation of antihydrogen atoms and the study of their quantum level distributions, as well as the exploration of antiprotonic atoms production.

On a broader scale, CIRCUS represents a novel kind of approach to managing experimental routines (and setups in general) with a focus on autonomy, which can be employed for a variety of different applications. In particular, experiments relying on precise synchronisation and coordination of subsystems handling individual tasks from different fields, reliable operation over many months, and flexible adaptations of the setup, such as those focused on atomic and quantum physics studies, can benefit from the introduction of this control system. The self-optimisation capabilities further render the system minimally sensible to external changes and very stable in its operation. 

Another experiment, PsICO (Positronium Inertial and Correlation Observation), has started implementing the CIRCUS to operate its apparatus: its main goal is to study the three-body entanglement properties of the three photons produced by the decay of ortho-Positronium, relating it to the initial spin state~\cite{bassColloquiumPositroniumPhysics2023,hiesmayrGenuineMultipartiteEntanglement2017}.

Both the hardware and software of the presented control system are available open-source to be adapted as needed for use in individual experiments, which is easily enabled by the modular and standardised library-based approach of the system's design.

%%%%%%%%%%%%%%%%%%%%%%%%%%%%%%%%%%%%%%%%%%%%%%%%%%%

\backmatter

\section*{Declarations}

\bmhead{Availability of data and materials}
\ \\
Data and materials are available under the conditions of the AEgIS collaboration Data Management Plan.
\bmhead{Competing interests}
\ \\
The authors declare that they have no competing interests.
\bmhead{Funding}
\ \\
The \Aegis Collaboration acknowledges the following funding agencies for their support:

This work is funded by the Research University – Excellence Initiative of Warsaw University of Technology via the strategic funds of the Priority Research Centre of High Energy Physics and Experimental Techniques, the IDUB POSTDOC programme, and by the Polish National Science Centre under agreements no. 2022/45/B/ST2/02029, and no. 2022/46/E/ST2/00255, and by the Polish Ministry of Education and Science under agreement no. 2022/WK/06.

This work has been sponsored by the Wolfgang Gentner Programme of the German Federal Ministry of Education and Research (grant no. 13E18CHA).

This work has been financed by the CERN Doctoral Student Programme, and by the Instituto Nazionale di Fisica Nucleare (INFN) - Sezione di Trento.

\bmhead{Authors' contributions}
\ \\
MV and SH have implemented, commissioned, and maintained the main pillars of the control system: MV has developed the TALOS LabVIEW\texttrademark\ framework and the slow-control elements for the integration of all experimental subsystems; SH has built up the Sinara electronics system and the ARTIQ library structure for the fast control and synchronisation of the experiment. MV and SH have also been part of the core team running the experiment during the antiproton beam times. MV and SH are the major contributors in writing the manuscript.

RC has devised the requirements of the control system and guided its implementation with his experience. RC has also been part of the core team running the experiment during the antiproton beam times and lead the corresponding data analysis.

JZ has designed and implemented core components of the TALOS infrastructure for a reliable operation of the experiment and contributed to the data taking during the antiproton beam times.

GK, GK, and DN have introduced the collaboration to the ARTIQ/Sinara portfolio and developed the high-voltage amplifier units.

TR has built up the ALPACA framework used for direct data analysis and for the integration of the self-optimisation capabilities of the system and contributed significantly to the data analysis.

BR has contributed to the ARTIQ library structure and the data taking with regard to the positronium and laser routines and the feedback loop used for the parameter optimisation.

FP has developed and maintained the \Aegis data acquisition system.

All authors revised and approved the final manuscript.

\bmhead{Acknowledgements}
\ \\
The \Aegis Collaboration would like to thank the CERN accelerator and decelerator teams for the outstanding performance of the AD--ELENA complex. \\
We also thank Dr. Simone Stracka for the inspiring discussions and Miko\l{}aj Sowi\'nski for his ARTIQ support.

\begin{appendices}

%%%%%%%%%%%%%%%%%%%%%%%%%%%%%%%%%%%%%%%%%%%%%%%%%

% The AEgIS Collaboration

%%%%%%%%%%%%%%%%%%%%%%%%%%%%%%%%%%%%%%%%%%%%%%%%%

%%%%%%%%%%%%%%%%%%%%%%%%%%%%%%%%%%%%%%%%%%%%%%%%%

% Calibration appendix

%%%%%%%%%%%%%%%%%%%%%%%%%%%%%%%%%%%%%%%%%%%%%%%%%

\section{Calibration of the voltage amplifiers}
\label{appendix_calibration}

\begin{figure}[H]
\begin{center}
\includegraphics[width = 0.8 \textwidth] {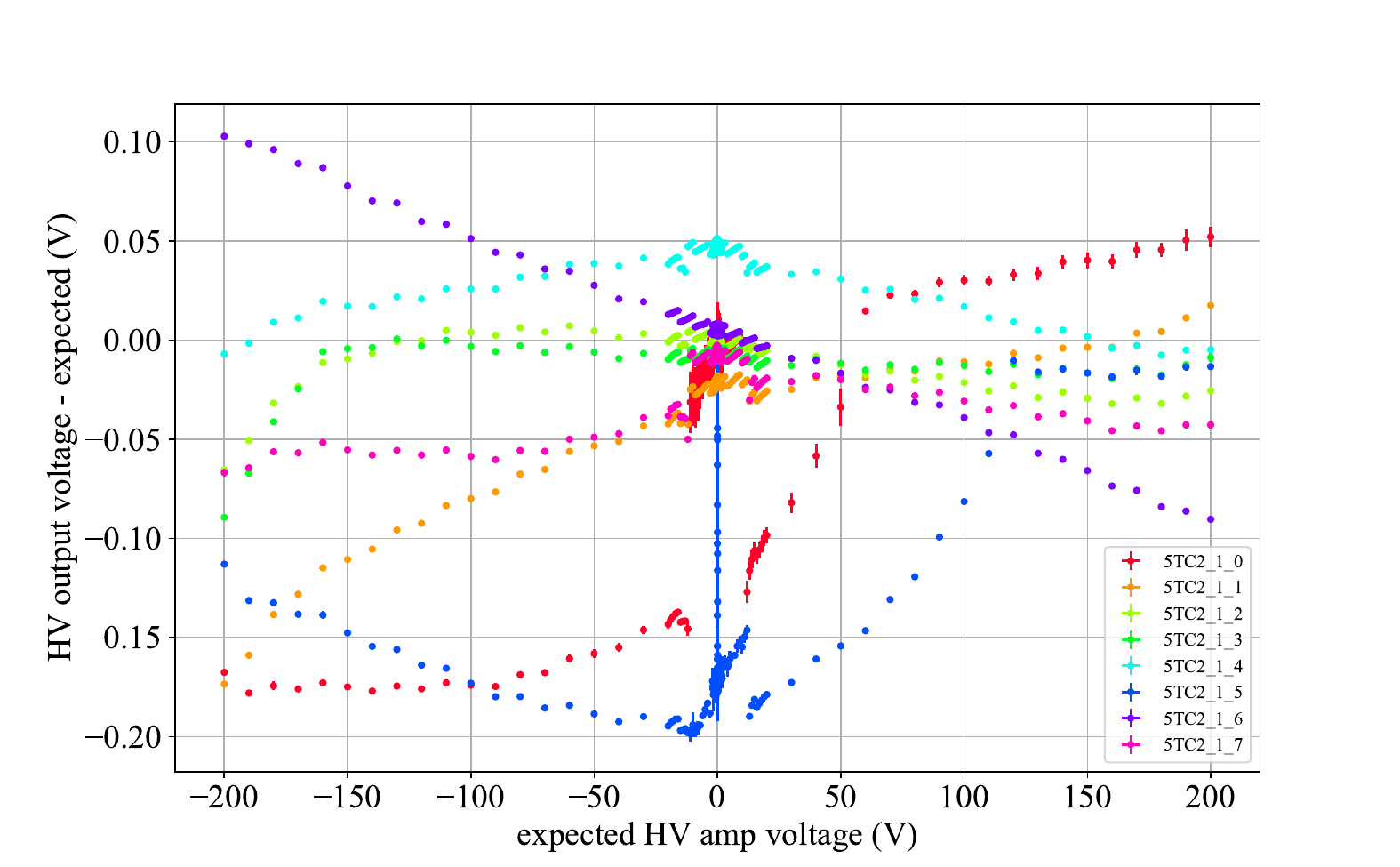} 
\includegraphics[width = 0.8 \textwidth] {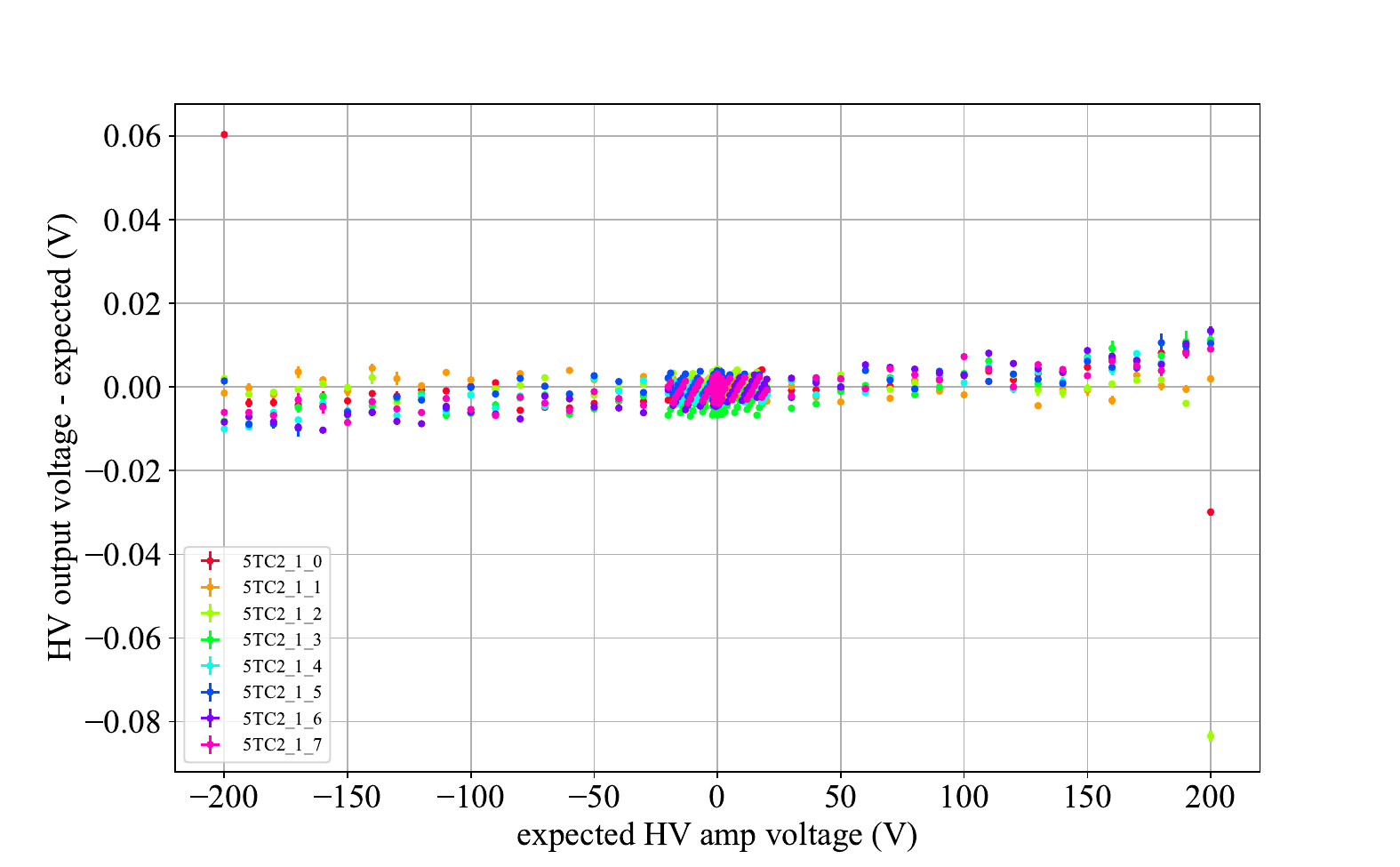}
\caption{Difference between the desired voltage on the amplifier channels and the measured output voltage versus the expected voltage before (top) and after (bottom) the amplifier calibration for all eight amplifier channels of one example board. The legend shown in the right plot is valid for both and identifies the channel numbers of the given board.}
\label{figure_amp_calibration}
\end{center}
\end{figure}

The calibration of the amplifier channels providing the electrode voltages to the \Aegis traps is based on a scan from the minimum to the maximum of the range of possible Fastino channel voltages (\SI{-10}{\volt} to +\SI{10}{\volt}, which corresponds to \SI{-200}{\volt} to +\SI{200}{\volt} on the amplifier outputs). As a compromise between statistical precision and timing efficiency, a step size of 327 machine units of the Fastino (approximately \SI{0.1}{\volt} depending on the exact configuration of each individual channel) is chosen for the scanning measurements. At every voltage step, five measurement iterations are performed for every channel to be calibrated, where the actually produced voltage on the amplifier output is registered by a multimeter and read to a calibration file in JSON format.
\\
The software calibration routine for all channels is then done at the same time: the data is fitted with a linear function from lowest to highest setting and the slope and offset are saved as calibration parameters for every channel individually. These parameters are imported into the corresponding ARTIQ library and directly applied as correction factors when setting a voltage on one of the trap electrodes from software.
\\
A verification measurement for each channel is executed after waiting for an arbitrary amount of time, thus excluding a systematic influence from environmental conditions. For these measurements, a different scan through the range of Fastino voltages is performed, directly applying the calibration correction. The produced voltages are read out in the same way and the data is analysed to verify the minimisation of the differences between desired and produced voltage by the calibration for all channels.
\\
Fig.~\ref{figure_amp_calibration} shows the result of this verification measurement before and after calibration for one of the amplifier boards, taken as example. 
\\
The absolute voltage accuracy reached after calibration is significantly improved, reaching the mV level on all channels, rendering it comparable to the 16-bit, i.e. \SI{6}{\milli\volt}, precision of the Fastino settings. The precise reachable minimum and maximum voltage depends on the individual internal configuration of the Fastino channel and causes larger deviations in either positive or negative direction when pushing to the very boundaries. However, the reachable value is in no case further away than 0.1\,V from the extremes of $\pm$\SI{200}{\volt}, which suffices for the purposes of \Aegis{}, as voltages beyond $\pm$\SI{190}{\volt} are never required for the application of the trap potentials. The clustering of data points at low absolute voltage values stems from the procedure of adapting the step size to the scanned voltage range for the verification measurements; the internal step structure is a consequence of the 16-bit precision of the voltage settings. The large fluctuations and resulting error bars for low voltage settings of some channels were caused by an intrinsic condition of the hardware, which has since been fixed by the mounting of additional capacitors in the amplifier circuits.

\end{appendices}


\bibliography{main.bib}

\end{document}